\documentclass[aps,pra,twocolumn,amsmath,amssymb,floats,floatfix,showpacs]{revtex4}

\usepackage{graphicx}
\usepackage{dcolumn}

\begin{document}

\title{Open system dynamics with non-Markovian quantum jumps} 

\author{J. Piilo}
\email{jyrki.piilo@utu.fi}
\affiliation{
Department of Physics and Astronomy, University of Turku, 
FI-20014 Turun yliopisto, Finland
}

\author{K. H\"ark\"onen}
\affiliation{
Department of Physics and Astronomy, University of Turku, 
FI-20014 Turun yliopisto, Finland
}

\author{S. Maniscalco}
\affiliation{
Department of Physics and Astronomy, University of Turku, 
FI-20014 Turun yliopisto, Finland
}

\author{K.-A. Suominen}
\affiliation{
Department of Physics and Astronomy, University of Turku, 
FI-20014 Turun yliopisto, Finland
}

\date{\today}

\begin{abstract}
We discuss in detail how non-Markovian open system dynamics can be described in terms of quantum
jumps [J. Piilo {\it et al.}, Phys.~Rev.~Lett.~{\bf 100}, 180402 (2008)].
Our results demonstrate that  it is possible to have a jump description contained in the physical Hilbert space of the reduced system. The developed non-Markovian quantum jump (NMQJ) approach is a generalization of the Markovian Monte Carlo Wave Function (MCWF) method into the
non-Markovian regime. 
The method conserves both the probabilities in the density matrix and the norms of the state vectors exactly,
and sheds new light on non-Markovian dynamics. The dynamics of the pure state ensemble illustrates how local-in-time master equation can describe memory effects and how the current state of the system carries information on its earlier state. Our approach solves the problem of negative jump probabilities of the Markovian MCWF method in the non-Markovian regime by defining the corresponding jump process with positive probability. 
The results demonstrate that in the theoretical description of non-Markovian open systems, there occurs quantum jumps which recreate seemingly lost superpositions due to the memory.

\end{abstract}


\maketitle

\section{Introduction}
The theory of open quantum systems describes the dynamics of a system of interest interacting with its environment~\cite{Breuer2002}.
The system-environment interaction leads to non-unitary reduced system dynamics and the system state is described by a density matrix instead of a single state vector used for closed systems. Generally, the density matrix evolution is governed by a master equation whose unitary part contains the dynamics as given by the system Hamiltonian and the non-unitary dissipator describes the effects that the environment has on the system. 

The presence of the environment leads to decoherence which is harmful for practical applications like quantum information processing~\cite{Stenholm2005}. On the other hand, decoherence has a role in open fundamental problems of quantum physics such as 
quantum to classical transition~\cite{Zurek}.
Often, the environment is seen to have unavoidable effects on the system dynamics. However, the recently developed ability to control quantum systems and the implementation of reservoir engineering techniques are revising the role of the environment~\cite{engineerNIST,control,Zoller}. This may lead to new ways to control the system of interest indirectly via the control of the system--reservoir interaction and the properties of the environment.

In memoryless Markovian open systems, the environment acts as a sink for the system information.
Due to the system-reservoir interaction, the system of interest loses information on its state into the environment, and this lost information does not play any further role in the system dynamics. However, if the environment has a non-trivial structure, then the seemingly lost information can return to the system at a later time leading to non-Markovian dynamics with memory. This memory effect is the essence of non-Markovian dynamics.

Non-Markovian systems appear in many branches of physics, such as quantum optics~\cite{Breuer2002,Gardiner96a, Lambro},
solid state physics~\cite{SS}, quantum chemistry~\cite{QC}, and quantum information processing~\cite{QIP}. Recently, non-Markovian features has also been exploited in the context of biomolecules where the environment consists of protein solvents~\cite{Thorwart08}.
However,  the elusive nature of non-Markovian dynamics makes it often difficult to obtain insight
into microscopical physical processes governing the time evolution. At the same time the complex mathematical structure of the
non-Markovian models prevents generally to solve 
the dynamics of the system of interest. Hence,  new ways to describe  non-Markovianity and new methods to solve non-Markovian dynamics are highly desired.

The density matrix can also be seen as a collection, or ensemble, of state vectors. 
Then, the interaction between the system and the reservoir removes the  precise information about the specific state vector to describe the system state.
Instead, the state of the open system is associated with an ensemble of state vectors where each state vector has a certain (classical) probability of appearance.
This view has led to the development of Monte Carlo simulation methods for Markovian~\cite{DCM1992,Dum92a,Carmichael, Plenio98,Gisin,Percival} and non-Markovian~\cite{Imamoglu,Garraway1997,Breuer99,Gambetta2004,BreuerGen,Piilo08,Strunz1999} open systems. 
In these methods, the time evolution of each state vector in the ensemble contains a stochastic element which can be discontinuous
(quantum jump)~\cite{DCM1992,Dum92a,Carmichael,Imamoglu,Garraway1997,Breuer99,Gambetta2004,BreuerGen,Piilo08} or continuous
(quantum state diffusion)~\cite{Gisin,Percival,Strunz1999}.

One of the most common methods to treat Markovian dynamics is the Monte Carlo wave function (MCWF) method which exploits quantum
jumps~\cite{DCM1992}. However, a generalization of this Markovian method to non-Markovian regime has turned out to be a challenging problem.
The central obstacle has been the appearance of negative quantum jump probabilities due to the temporarily negative decay rates of non-Markovian dynamics.
Earlier approaches to this problem exploit auxiliary extensions
of the Hilbert space of the system~\cite{Imamoglu,Garraway1997,Breuer99,BreuerGen} or exploit the state of the total system~\cite{Gambetta2004}.

We have recently shown that the jump-like unravelling of non-Markovian master equations is possible within the Hilbert space of the system,  and hence the auxiliary extension of the system Hilbert space is not necessarily needed~\cite{Piilo08}.
The key feature of the developed non-Markovian quantum jump (NMQJ) method is the notion that,
when the decay rates appearing in the master equation become negative,
the direction of the information flow between the system and the reservoir gets reversed.
During the initial positive decay region, the information flows from the system to the environment, while during the negative decay the system
may regain some of the 
information it lost earlier.  In terms of quantum jumps this means that the seemingly lost superpositions in the ensemble can be restored.
This leads to new insight into the concept of memory, which is the central ingredient of non-Markovian dynamics.
We also describe in detail the positive and negative factors affecting the numerical performance of the method. The ultimate limit for the numerical performance is given by the effective ensemble size $N_{\rm eff}$ (Sec.~\ref{Sec:Num}) since the method needs to evolve simultaneously $N_{\rm eff}$ state vectors.

Our results help to explain why local-in-time master equations \cite{Breuer2002,Andersson} can indeed describe systems with memory and the results also show the presence of some counterintuitive features of non-Markovian dynamics. In this regime, the rate of the process is proportional to the target state, instead of the source state, and hence challenges the classical view. We show here two different  proofs of the equivalence between the algorithm and the master equation, discuss in detail how the method works, and apply it to multi-level atom schemes. 
Recently, the existence of a measurement scheme interpretation of non-Markovian dynamics has been actively discussed~\cite{Diosi08,Gambetta08}. Our results align along the results of Ref.~\cite{Gambetta08}. We discuss this and other insight provided by the NMQJ method in detail.

We have organized the paper in the following way. Sec.~\ref{MCWF} describes briefly the Markovian MCWF method and sets the scene for its non-Markovian generalization which is presented in Sec.~\ref{NMQJ}. We then present several examples on the use of the NMQJ method in Sec.~\ref{Exs} and discuss the insight provided by the method in Sec.~\ref{Discu}. Finally, Sec.~\ref{Conclu} concludes the paper.

\section{Markovian Monte Carlo wave function method}\label{MCWF}

Our non-Markovian quantum jump method generalizes the  MCWF method~\cite{DCM1992} into the non-Markovian regime. The algorithms and the proof of correspondence with the master equation for the two methods are very similar. The essential difference is the form of the jump operators and jump probabilities.
We present first the central ingredients of the Markovian MCWF method and illustrate the problems that prevents its use for non-Markovian systems.

\subsection{The algorithm and equivalence with master equation}

The MCWF method is probably the most commonly used Monte Carlo method to treat Markovian open systems 
whose dynamics is governed by the master equation in the Lindblad form~\cite{DCM1992,Gorini}
\begin{eqnarray}
\dot{\rho} (t) &=& \frac{1}{\imath\hbar} \left[ H_S, \rho (t) \right]   + \sum_j\Gamma_j C_j \rho(t)C_j^{\dag}
\nonumber \\
&-& 
\frac{1}{2}\sum_j\Gamma_j \left\{C_j^{\dag} C_j, \rho(t)\right\}.
\label{Eq:Mark}
\end{eqnarray}
Here, $\rho$ is the density matrix of the reduced system, $H_S$ the hermitian system Hamiltonian, $\Gamma_j$ is the positive and constant decay rate to decay channel $j$, and $C_j$ are the Lindblad (jump) operators describing the effects of the environment on the reduced system.

To unravel the master equation (\ref{Eq:Mark}), MCWF method generates an ensemble of stochastic state vector realizations whose deterministic and continuous time evolution is interrupted by randomly occurring discontinuous quantum jumps. The average over the ensemble of stochastic realizations gives the properties of the reduced system at any given moment of time. A generic way to write the density matrix in terms of the ensemble is 
\begin{equation}
\label{Eq:Rho}
\rho(t) = \sum_\alpha \frac{N_\alpha(t)}{N} |\psi_\alpha(t)\rangle \langle \psi_\alpha(t)|,
\end{equation}
where $N_\alpha(t)$ is the number of ensemble
members in the state $|\psi_\alpha(t)\rangle$ at time $t$
and $N$ is the total number of state vectors in the ensemble (ensemble size).

The method proceeds in discrete time steps $\delta t$, and we consider one step that takes us from time $t$ to $t+\delta t$.
During this time step, a given state vector $|\psi_{\alpha}(t)\rangle$ evolves either in a deterministic way
or performs a randomly occurring quantum jump. The deterministic evolution is given by the non-Hermitian Hamiltonian 
\begin{equation}
H= H_S -\frac{i\hbar}{2}\sum_j\Gamma_jC_j^{\dagger}C_j.
\label{Eq:H}
\end{equation}
The essential feature here is the second term on the r.h.s., which is constructed from the jump operators that appear in the master equation (\ref{Eq:Mark}). This term reduces, in the Markovian case, the occupation probability of the states which decay. 
The deterministic time-evolution by the Hamiltonian (\ref{Eq:H}) leads, for small enough time step $\delta t$, to the state
\begin{equation}
\label{Eq:Phi}
| \phi_\alpha(t+\delta t)\rangle = 
\left(1-\frac{iH\delta t}{\hbar}\right)  |\psi_\alpha(t)\rangle.
\end{equation}
Before the next time step, this state is renormalized and the time evolution of $|\psi_{\alpha}\rangle$ is
\begin{equation}
\label{Eq:Det}
|\psi_{\alpha}(t)\rangle \rightarrow |\psi_\alpha(t+\delta t)\rangle = \frac{ |\phi_\alpha(t+\delta t)\rangle} {|||\phi_\alpha(t+\delta t)\rangle ||}.
\end{equation}

If, instead of the deterministic evolution, a quantum jump to channel $j$ occurs, the state vector changes in a discontinuous way
\begin{equation}
\label{Eq:Jump}
|\psi_\alpha(t)\rangle  \rightarrow |\psi_\alpha(t+\delta t)\rangle =\frac{C_{j} |\psi_\alpha(t)\rangle} {||C_j| \psi_\alpha(t)\rangle||}.
\end{equation}
The probability $p_\alpha^{j}$ for a state vector $|\psi_{\alpha}\rangle$  to have a quantum jump to channel $j$ is
directly proportional to the corresponding decay rate $\Gamma_j$, the time step size $\delta t$ and the occupation probability of the decaying state
\begin{equation}
\label{Eq:pj}
p_\alpha^{j}(t)=\Gamma_{j}\delta t \langle \psi_\alpha(t) | C_j^{\dagger}C_{j}|\psi_\alpha(t)\rangle.
\end{equation}

The choice between the deterministic and jump evolutions, Eqs.~(\ref{Eq:Det}) and (\ref{Eq:Jump}) respectively, is done by 
comparing  a generated random number $\xi$ to the total jump probability $p_{\alpha}$. This is the sum over channel specific probabilities $p_{\alpha}^j$
\begin{equation}
p_{\alpha}=\sum_jp_{\alpha}^j,
\end{equation}
and has a direct relation to the norm of $| \phi_\alpha(t+\delta t)\rangle$: $1-p_{\alpha}=|||\phi_\alpha(t+\delta t)\rangle ||^2$.

By calculating the average evolution $\overline{\sigma_{\alpha}} $ of $|\psi_{\alpha}(t)\rangle$ over the deterministic and jump paths one obtains
\begin{eqnarray}
\label{Eq:Ave}
\overline{\sigma_{\alpha}(t+\delta t)}
&=&
(1-p_{\alpha}) \frac{| \phi_{\alpha}(t+\delta t)\rangle\langle \phi_{\alpha}(t+\delta t) |}{1-p_{\alpha}}
\nonumber \\
&+&
\sum_j p_{\alpha}^j \frac{C_j| \psi_{\alpha}(t)\rangle\ \langle \psi_{\alpha}(t) |C_j^{\dagger}}{\langle \psi_{\alpha}(t) | C_j^{\dagger}C_j|\psi_{\alpha}(t)\rangle}.
\label{Eq:SimuMark}
\end{eqnarray}
Here, $(1-p_{\alpha})$ is the no-jump probability which weights the deterministic evolution and jump probabilities
$p_{\alpha}^j$ weight the corresponding jump paths. 

By inserting Eqs.~(\ref{Eq:Phi}) and (\ref{Eq:pj}) into the
Eq.~(\ref{Eq:Ave}) and rearranging the terms, one obtains after straightforward calculation the master equation (\ref{Eq:Mark}) for
state vector $|\psi_{\alpha}(t)\rangle$. Taking a further step by considering the average over the whole ensemble,
\begin{equation}
\label{Eq:MCAve}
\overline{\sigma(t+\delta t)}=\sum_\alpha \frac{N_{\alpha}}{N}\overline{\sigma_{\alpha}(t+\delta t)},
\end{equation}
it is straightforward to see that the master equation (\ref{Eq:Mark}) and the MCWF method result given by Eq.~(\ref{Eq:MCAve}) match,
and the two approaches are indeed equivalent descriptions of the Markovian open system dynamics.

\subsection{Why the MCWF does not work for non-Markovian systems?}

In Markovian systems, the decay and decoherence processes occur at constant positive rates [c.f.~Eq.~(\ref{Eq:Mark})]. This indicates constant flow of information from the system to the environment before the steady state is reached. For non-Markovian systems, the decay rates are time-dependent and may acquire temporarily negative values (to be described in detail in the next Section). During the initial period of  positive time dependent decay, 
the rate of the information flow changes but the direction of the flow remains constant, i.e., from the system to the environment.
When the decay rate becomes negative, the direction of the information flow is reversed and the reduced system, due to the non-Markovian memory, begins to recall the information that was lost earlier. 

In the MCWF method, the quantum jump probability is directly proportional to the decay rate [c.f.~Eq.~(\ref{Eq:pj})] which acquires negative values in the non-Markovian case. As a consequence of these two facts, a quantum jump has negative probability to occur while the deterministic evolution has larger than $1$ probability. Therefor, it is impossible to make a decision between these two alternatives and as a consequence, MCWF method can not be used to describe non-Markovian dynamics.

Earlier attempts to solve this problem exploit usually the idea that non-Markovian dynamics can be converted to Markovian one by extending the Hilbert space of the system~\cite{Imamoglu,Garraway1997,Breuer99,BreuerGen}. This may come with a cost for computational efficiency and may also prevent obtaining insight into non-Markovian dynamics. This also leaves open a fundamental question: Is there a corresponding 
jump process  in the Hilbert space of the system which has a positive probability?

\section{Non-Markovian quantum jumps}\label{NMQJ}

Our starting point is the general local-in-time non-Markovian master equation ~\cite{Breuer2002,BreuerGen}
\begin{eqnarray}
\dot{\rho}(t) &=&  \frac{1}{i\hbar} \left[ H_S, \rho(t)\right] + 
\sum_j\Delta_j(t) C_j(t) \rho(t)C_j^{\dag}(t) \nonumber \\
& -&\frac{1}{2}\sum_j\Delta_j(t)\left\{\rho(t),C_j^{\dag}(t) C_j(t) \right\}.
 \label{Eq:MNM}
\end{eqnarray}
The difference, when comparing to the Markovian master equation~(\ref{Eq:Mark}), is that the decay rates $\Delta_j(t)$ depend on time and may acquire negative values. In the most general case the Lindblad operators $ C_j(t)$ may also depend on time.

\subsection{Special case: Non-Markovian time scale is the shortest one\label{sec:simpleCase}}

Before going to the general solution in the next subsection, we first describe the method for the simple case in which
the non-Markovian time-scale is the fastest one, which is most often the case. This allows to introduce the NMQJ method in a way that is conceptually rather straightforward.
With this approximation, the state vectors do not have time to evolve due to the system Hamiltonian $H_S$ on the time scale of non-Markovian dynamics.  Consider now a non-Markovian system where the decay rates oscillate between positive and negative values before reaching a constant Markovian value.
For the sake of simplicity, we assume here first that all the decay channels take negative values simultaneously. 
At the end of the first positive decay period,
the initial pure state has evolved to a mixed state which can be described in terms of the jump paths,
unravelled by the MCWF method in the positive region, as
\begin{eqnarray}
\label{Eq:Rho1}
\rho(t) &=& \frac{N_0}{N} |\psi_0(t)\rangle \langle \psi_0(t) | + \sum_j   \frac{N_j}{N} |\psi_j\rangle \langle \psi_j | 
\nonumber \\
&+&
\sum_{j,k}   \frac{N_{j,k}}{N} |\psi_{j,k}\rangle \langle \psi_{j,k} | + ...
\end{eqnarray}
Here, $|\psi_0(t)\rangle$ is the deterministic evolution from the initial state $|\psi_0(0)\rangle$ without jumps and 
$|\psi_j\rangle$ describe the ensemble members that have performed one jump to channel $j$, such that $|\psi_j\rangle= C_j |\psi_0\rangle / || C_j  |\psi_0\rangle||$. In the next term, $|\psi_{j,k}\rangle$ correspond to members who have performed first a jump to channel $j$ and then, furthermore, a second jump to channel $k$, so that $|\psi_{j,k}\rangle= C_kC_j  |\psi_0\rangle / || C_kC_j  |\psi_0\rangle||$. The rest of the terms go in the corresponding way.
$N_0$, $N_j$ and $N_{j,k}$ are the corresponding numbers of the ensemble members.


\begin{figure}[tb]
\includegraphics{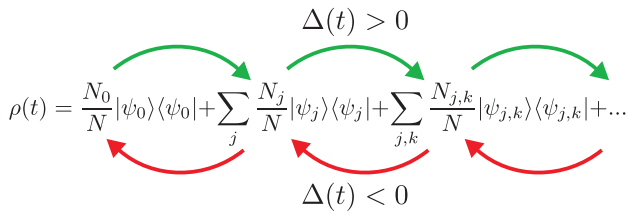}
\caption{\label{Fig:probFlow} 
(Color online) Initially all the $N$ ensemble members share the same initial state $|\psi_0\rangle$, i.e., $N_0(0)=N$. Quantum jumps during the positive decay rate (arrows to the right) spread the ensemble members to a wider set of different states. On the contrary, the non-Markovian quantum jumps during the negative decay rate (arrows to the left) transfer ensemble members always to states which already exist in the ensemble.
}
\end{figure}

The central question is now how the ensemble (\ref{Eq:Rho1}) is evolved so that the result matches the master equation (\ref{Eq:MNM}). 
The sign change of the decay rate indicates the reversal of the information flow between the system and the environment, so that for negative decay 
the system partially recovers the information that it lost earlier. This restoration of lost information is the essence of the non-Markovian memory. In other words, the decoherence that occurred in the preceding  positive decay region, turns to re-coherence in the negative decay region, i.e., the earlier effects of decoherence get partially cancelled.

This leads to the idea that non-Markovian quantum jumps, taking place in the negative decay region, cancel the effect of the jumps that appeared earlier in the positive decay region destroying quantum superpositions. Reverse quantum jumps during negative decay are thus expected to counteract 
prior positive decay jumps.
This means that in the expansion (\ref{Eq:Rho1}), state $|\psi_j\rangle$ jumps back to the state $|\psi_0\rangle$, state $|\psi_{j,k}\rangle$ jumps to the state $|\psi_j\rangle$, and so on. The direction of the probability flow gets reversed for negative decay region as
illustrated in Fig.~\ref{Fig:probFlow}. The corresponding non-Markovian quantum jump operators are
\begin{eqnarray}
\label{Eq:JOps1}
D_{j\rightarrow 0} &=& |\psi_0(t)\rangle \langle \psi_{j}|, \nonumber \\
D_{j,k\rightarrow j} &=& |\psi_j \rangle \langle \psi_{j,k}|,
\end{eqnarray}
and so on. The probabilities for the jumps to occur are
\begin{eqnarray}
\label{Eq:JProbs1}
P_{j\rightarrow 0} &=&
\frac{N_0 \delta t |\Delta_j| \langle \psi_0(t) | C_j^{\dagger} C_j |\psi_0(t)\rangle} {N_{j}},
\nonumber \\
P_{j,k\rightarrow j} &=&
\frac{N_j \delta t |\Delta_j| \langle \psi_j | C_j^{\dagger} C_j |\psi_j \rangle}{N_{j,k}}.
\end{eqnarray}
Equations (\ref{Eq:JOps1}) and (\ref{Eq:JProbs1}) demonstrate that the probability for reversing a jump for one particular channel is given by the portion of ensemble members that have not yet jumped in that channel. The numerator gives the total jump probability in the ensemble 
which is distributed equally
to those ensemble members which can perform the jumps. By doing the reversed jump according to Eq.~(\ref{Eq:JOps1}), the discontinuous history of the ensemble member is preserved. This means that when we are reversing a jump, we are not erasing the past.

To prove that the algorithm matches with the master equation, we follow very closely the proof of the MCWF method~\cite{DCM1992}. The basic idea is to average over the deterministic and jump paths 
in order to obtain an equation of motion for the reduced density matrix.
Evolving the 
ensemble (\ref{Eq:Rho1}) over time step $\delta t$, gives
\begin{eqnarray}
\overline{\sigma(t+\delta t)} &=& \Theta_0(t) + 
\sum_j \frac{N_j}{N}
\left[
\Theta_j(t)+
\Theta_{j\rightarrow 0}(t) \right]
\nonumber \\
&+&
\sum_{j,k}
\frac{N_{j,k}}{N}
\left[\Theta_{j,k}(t) + \Theta_{j,k\rightarrow j}(t) \right]+ ...
\label{Eq:Ave1}
\end{eqnarray}
Here, $\Theta_0(t)$ is the contribution arising from the deterministic evolution between times $0$ and $t$.
$\Theta_j(t)$ is the contribution of the ensemble members who jumped earlier once to channel $j$ and the jump is not cancelled at the 
current point of time.
In $\Theta_{j\rightarrow 0}(t)$, there has been one jump to channel $j$ and which gets cancelled at the current point of time.
The rest of the terms arise correspondingly. It is worth noting that it is not possible to cancel something which never happened. Hence there are no jumps which can be cancelled from $\Theta_0(t)$ part.
Taking into account for the appropriate weights and keeping in mind the jump operators and probabilities from
Eqs.~(\ref{Eq:JOps1}) and (\ref{Eq:JProbs1}), these terms can be written explicitly
\begin{eqnarray}
\label{Eq:Thetas}
\Theta_0 &=&
\frac{| \phi_0(t+\delta t)\rangle \langle \phi_0(t+\delta t) |}{1+n_0},
\nonumber \\
\Theta_j(t) &=&
(1-P_{j\rightarrow 0}) \frac{| \phi_{j}(t+\delta t)\rangle \langle \phi_{j}(t+\delta t) |}{1+n_{j}},
\nonumber \\
\Theta_{j\rightarrow 0}(t) &=&
P_{j\rightarrow 0} D_{j\rightarrow 0} |\psi_{j}(t)\rangle \langle \psi_{j}(t) |
D_{j\rightarrow 0}^{\dagger}.
\end{eqnarray}
Here, the time evolved deterministic states are
\begin{eqnarray}
\label{Eq:Det1}
| \phi_0(t+\delta t)\rangle &=&
(1-\frac{iH_S\delta t}{\hbar} +\sum_m\frac{|\Delta_m|\delta t}{2} C_m^{\dagger}C_m) |\psi_0(t)\rangle,
\nonumber \\
| \phi_{j}(t+\delta t)\rangle &=&
(1-\frac{iH_S\delta t}{\hbar} +\sum_m\frac{|\Delta_m|\delta t}{2} C_m^{\dagger}C_m) |\psi_j(t)\rangle,
\nonumber \\
\end{eqnarray}
and their normalization factors are
\begin{eqnarray}
\label{Eq:Norm1}
n_0&=&\sum_m\delta t |\Delta_m(t)| \langle \psi_0(t) | C_m^{\dagger} C_m |\psi_0(t)\rangle,
\nonumber \\
n_j&=&\sum_m\delta t |\Delta_m(t)| \langle \psi_j(t) | C_m^{\dagger} C_m |\psi_j(t)\rangle.
\end{eqnarray}
All the rest of the terms follow correspondingly.
Using Eqs.~(\ref{Eq:Det1}) and (\ref{Eq:Norm1}) in Eq.~(\ref{Eq:Thetas}) and inserting the results into
Eq.~(\ref{Eq:Ave1}) gives the master equation (\ref{Eq:MNM}). 

In a multi-channel system, positive and negative channels may appear simultaneously. The description above contains all the negative channels while the positive channels evolve according to the MCWF method. Hence, the match between the positive channel dynamics with the master equation can be proven along the MCWF proof. For the sake of simplicity, we leave the detailed description of simultaneous positive and negative channels to the general treatment presented in the next subsection.

\subsection{General case}

The simplified case presented in the previous section \ref{sec:simpleCase} is now generalized. The simple treatment fails in a general case, because it assumes that the jump history can be unambiguously reconstructed for each state in the decomposition \eqref{Eq:Rho1}. In general, starting from $|\psi_0\rangle\langle \psi_0 |$, many different combinations of jumps may lead to identical contribution  $|\psi_\alpha\rangle\langle \psi_\alpha |$, and all these states should be counted together to form $N_\alpha$. 

As in the Markovian case, we write the density matrix in the most generic way
\begin{equation}
\label{Eq:Rho2}
\rho(t) = \sum_\alpha \frac{N_\alpha(t)}{N} |\psi_\alpha(t)\rangle \langle \psi_\alpha(t)|.
\end{equation}
The positive and negative decay channels are noted with $j_+$ and $j_-$, respectively, while the corresponding decay rates are 
$\Delta_{j_+}(t)>0$ and $\Delta_{j_-}(t)<0$. With this notation the master equation (\ref{Eq:MNM}) can be written as
\begin{widetext}
\begin{eqnarray}
\dot{\rho}(t) &=&  \frac{1}{i\hbar} \left[ H_S, \rho(t)\right] 
+\sum_{j_+}\Delta_{j_+}(t) \left[ C_{j_+}(t) \rho(t)C_{j_+}^{\dag}(t)
-\frac{1}{2}\left\{\rho(t),C_{j_+}^{\dag}(t) C_{j_+}(t) \right\}\right]
\nonumber \\
&-&
\sum_{j_-}|\Delta_{j_-}(t)| \left[ C_{j_-}(t) \rho(t)C_{j_-}^{\dag}(t)
-\frac{1}{2}\left\{\rho(t),C_{j_-}^{\dag}(t) C_{j_-}(t) \right\}\right]. \nonumber \\
 \label{Eq:MNM2}
\end{eqnarray}
\end{widetext}

The deterministic time evolution of the state vectors $|\psi_{\alpha}(t)\rangle$ occurs as before
\begin{equation}
\label{Eq:Det2}
|\psi_{\alpha}(t)\rangle \rightarrow |\psi_\alpha(t+\delta t)\rangle = \frac{ |\phi_\alpha(t+\delta t)\rangle} {|||\phi_\alpha(t+\delta t)\rangle ||},
\end{equation}
where the non-normalized state $ |\phi_\alpha(t+\delta t)\rangle$ has been obtained with the usual non-Hermitian Monte Carlo Hamiltonian.
For the sake of convenience, we write this Hamiltonian separating 
the positive and negative channels
\begin{eqnarray}
H &=& H_S-\frac{i\hbar}{2}\sum_{j_+}\Delta_{j_+}(t)C_{j_+}^{\dagger} (t) C_{j_+}(t)
\nonumber \\
&-& \frac{i\hbar}{2}\sum_{j_-}\Delta_{j_-}(t)C_{j_-}^{\dagger} (t) C_{j_-}(t).
\label{eq:deterministicEvolution}
\end{eqnarray}

The jump probabilities and the jumps for the positive channels $j_+$ follow the MCWF prescription, i.e.,
\begin{equation}
P_\alpha^{j_+}(t)=\Delta_{j_+} (t)\delta t \langle \psi_\alpha (t) | C_{j_+}^{\dagger}(t)C_{j_+}(t)|\psi_\alpha (t)\rangle,
\label{eq:positiveJumpProbability}
\end{equation}
and
\begin{equation}
\label{Eq:Jump2}
|\psi_\alpha(t)\rangle  \rightarrow |\psi_{\alpha'} (t+\delta t)\rangle =\frac{C_{j_+} |\psi_\alpha(t)\rangle} {||C_{j_+}| \psi_\alpha(t)\rangle||}, 
\end{equation}
correspondingly.

For negative channels $j_-$ the direction of the jump process gets reversed
\begin{equation}
\label{Eq:Jump3}
|\psi_{\alpha'} (t+\delta t)\rangle  \leftarrow |\psi_\alpha(t)\rangle =\frac{C_{j_-} |\psi_{\alpha'} (t)\rangle} {||C_{j_-}| \psi_{\alpha'} (t)\rangle||}.
\end{equation}
In other words, the
jump operator for negative channels takes the form
\begin{equation}
D_{\alpha\rightarrow \alpha'}^{j_-}(t)= |\psi_{\alpha'}(t)\rangle \langle \psi_{\alpha}(t)|,
\label{Eq:JOp}
\end{equation}
where the {\it source state} of the jump is $|\psi_{\alpha}(t)\rangle = C_{j_-}(t) |\psi_{\alpha'}(t)\rangle / ||C_{j_-}(t) |\psi_{\alpha'}(t)\rangle||$. The source and target state of the jump swap their role when the decay rate becomes negative.

This transition for a given state vector $|\psi_{\alpha}\rangle$ in the ensemble (\ref{Eq:Rho2}) occurs with probability
\begin{eqnarray}
P_{\alpha\rightarrow \alpha'}^{j_-}(t) = \frac{N_{\alpha'}(t)}  {N_{\alpha}(t)} |\Delta_{j_-}(t)| \delta t  \langle \psi_{\alpha'}(t) | C_{j_-}^{\dagger}(t) 
C_{j_-}(t) |\psi_{\alpha'}(t)\rangle.\nonumber \\
 \label{Eq:JProb2M}
\end{eqnarray}
Note that the probability of the non-Markovian jump is given by the target state $ |\psi_{\alpha'}\rangle$ of the jump along the term
$ \langle \psi_{\alpha'}(t) | C_{j_-}^{\dagger}(t) 
C_{j_-}(t) |\psi_{\alpha'}(t)\rangle$. Moreover, if there are no ensemble members in the target state, $N_{\alpha'}=0$, then the jump probability is equal to zero. 

The sign of the decay rate $\Delta_j(t)$ can be understood in the following way. First, when for a given channel $j$,
$\Delta_j(t)>0$,  the process goes as  $|\psi\rangle\rightarrow |\psi'\rangle =C_{j}|\psi)\rangle / ||C_{j} |\psi\rangle||$.
Later on, when the decay rate becomes negative,  $\Delta_j(t)<0$,  the direction of this process is reversed and the jump occurs to opposite direction $|\psi\rangle\leftarrow |\psi'\rangle$.

Generally, Eq.~(\ref{Eq:Jump3}) indicates that the explicit target state $|\psi_{\alpha'}(t)\rangle$ of the reverse jump for the source state  $|\psi_{\alpha}(t)\rangle$ is not necessarily unique. This means that the ensemble members in the state $|\psi_{\alpha}(t)\rangle$ can jump to different target states along Eq.~(\ref{Eq:Jump3}) whenever the corresponding jump probability is larger than zero. The major factor for the computational cost is defined by how many different types of states vectors are created during the positive decay region and the need to evolve them simultaneously due to their dependence in the negative decay region. This point is discussed more in Section \ref{Sec:Num}.

The proof of our NMQJ method follows again the same lines of the Markovian 
MCWF method~\cite{DCM1992} and given in the previous Section. By weighting the deterministic and jump paths over the time step $\delta t$ with the appropriate probabilities we obtain the master
equation (\ref{Eq:MNM}). 
Calculating the average $\overline{\sigma}$ of the evolution  of the ensemble (\ref{Eq:Rho2}) over
$\delta t$ gives 
\begin{widetext}
\begin{eqnarray}
\label{Eq:AlgoM}
\overline{\sigma(t+\delta t)} 
&=&
\sum_{\alpha}\frac{N_{\alpha}(t)}{N} 
\left[
\left(
1-\sum_{j_+}P_{\alpha}^{j_+}(t) -\sum_{j_-,\alpha'} P_{\alpha\rightarrow \alpha'}^{j_-}(t)\right)
\right.
\frac{| \phi_{\alpha}(t+\delta t)\rangle \langle \phi_{\alpha}(t+\delta t) |}{||| \phi_{\alpha}(t+\delta t)\rangle ||^2}
\nonumber \\
&+&
\sum_{j_+} P_\alpha^{j_+}(t)
\frac{C_{j_+}(t) |\psi_{\alpha}(t)\rangle \langle \psi_{\alpha}(t)| C_{j_+}^{\dagger}(t) }
{||C_{j_+}(t) |\psi_{\alpha}(t)\rangle||^2}
+
\left.
\sum_{j_-,\alpha'} P_{\alpha\rightarrow \alpha'}^{j_-}(t) D_{\alpha\rightarrow \alpha'}^{j_-}(t) |\psi_{\alpha}(t)\rangle \langle \psi_{\alpha}(t)| D_{\alpha\rightarrow {\alpha}'}^{j_- \dagger} (t)
\right].
\end{eqnarray}
\end{widetext}
Here, the summations $\alpha$ and $\alpha'$ run over the ensemble [c.f.~Eq.~(\ref{Eq:Rho2})], the summation over $j_+$ and $j_-$ cover the positive and negative channels, respectively. 
The first term on the r.h.s.,~in the summation over $\alpha$, is the product of the no-jump probability and the deterministic evolution of the state vector, the second and third terms describe the positive and negative channel jumps, respectively, with the corresponding probabilities. 

The details of the proof are presented in Appendix A and we describe here briefly the main features. Like in the Markovian MCWF case, the deterministic evolution gives the commutator and the anticommutator parts of the master equation. Moreover, the jump part of positive channels goes along MCWF giving the remaining "sandwich" term for positive channels $j_+$. After making the series expansion of the denominator of the deterministic part and keeping the terms to the first order in $\delta t$, we are left with the norm change term due to negative channels times the deterministic evolution, jump probability for negative channels times the deterministic evolution, and the jump term for negative channels. As shown in Appendix A, the first and last of these three cancel each other and the second one gives the "sandwich" term of the master equation (\ref{Eq:MNM2}) for negative channels. This completes the proof.

\section{Examples}\label{Exs}

In order to demonstrate the applicability of the NMQJ method we give now concrete examples. These examples also show how the method works at the level of single realizations in the ensemble. Our physical system of choice is an atom interacting with a Lorentzian structured reservoir, e.g., an atom interacting with a single mode of a leaky cavity. 

\begin{figure}[ht]
\includegraphics{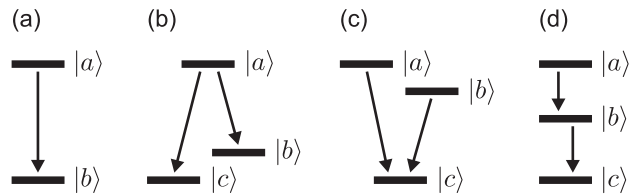}
\caption{\label{fig:examplesSchematic} Example cases introducing the level notations. (a) Jaynes-Cummings model, (b) $\Lambda$-system, (c) V-system, and (d) ladder-system. Only transitions expressed by arrows contribute since they reside close to the resonance frequency of the cavity.}
\end{figure}

The first example is a two-level atom interacting off-resonantly with the cavity field, also known as detuned Jaynes-Cummings model [c.f. Fig.~\ref{fig:examplesSchematic}~(a)]. We use this simple system to give a detailed walk-through description on how the NMQJ method is implemented in practice. The other examples deal with a three-level atom, another archtype of atomic systems, which holds two independent decay channels and also three different level geometries: $\Lambda$, V, and ladder-systems [c.f. Fig.~\ref{fig:examplesSchematic}~(b)--(d)]. For these cases we see how having simultaneously both a negative and a positive channel results in rich dynamics. 

\begin{figure}[tb]
\includegraphics{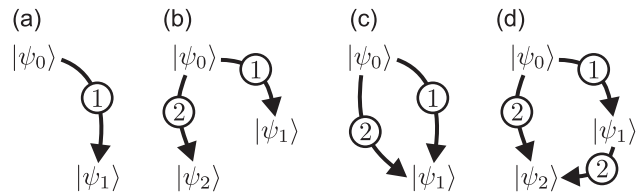}\label{Fig:Routes}
\caption{\label{fig:examplesEnsemble} Effective ensembles for the example cases. (a) Jaynes-Cummings model, (b) $\Lambda$-system, (c) V-system, and (d) ladder-system. The number in the arrow indicates the jump channel. The expressions for states $|\psi_\alpha\rangle$ are given in the text.}
\end{figure}

The structure of the effective ensemble, i.e., states $|\psi_\alpha \rangle$ and the way in which they connect by different jump channels, is shown in Fig.~\ref{fig:examplesEnsemble} for each example case, respectively. This illustrates, how physically identical states can be reached by different combinations of jumps in the V and ladder-systems.

From the NMQJ method's point of view the details of the actual physical system and the variety of approximations during the derivations are irrelevant as long as the master equation is in the desired general form, given by Eq.~\eqref{Eq:MNM}. Moreover, just to highlight this feature, we illustrate explicitly how the NMQJ method follows the formal mathematical solution of the given master equation as far as the solution is physically consistent, i.e., the density matrix remains positive. If the solution fails to be positive at some point, it obviously means that some of the approximations made while deriving the master equation of the reduced system are not valid.

\subsection{Derivation of the non-Markovian local-in-time Master equation}

To give an idea of how non-Markovian local-in-time master equations can be derived microscopically and of the explicit form of the time-dependent decay rates, we give a brief sketch of the derivation for the example system in hand.

The system Hamiltonian of a multi-level atom is
\begin{equation}
H_S = \sum_i \hbar \omega_i |i\rangle \langle i|.
\end{equation}
Similarly, the self-Hamiltonian for the electromagnetic field constituting the environment is
\begin{equation}
H_\textrm{env} = \sum_k \hbar \nu_k a_k^\dagger a_k.
\end{equation}
The dipole interaction between the system and its environment is described by an interaction Hamiltonian
\begin{equation}
H_\textrm{int} = - \mathbf{D} \cdot \mathbf{E},
\end{equation}
where $\mathbf{D}=q \mathbf{r}$ is the dipole moment operator and
$\mathbf{E}$ the quantized electromagnetic field.
Within the second order time-convolutionless (TCL) approach~\cite{Breuer2002} and after performing the secular approximation, the jump channels are categorized by atomic transition frequencies, or Bohr frequencies, $\omega$, such that the Lindblad operators are
\begin{equation}
C_{\omega} = \sum_{\substack{i,j:\\\omega_{j}-\omega_{i} = \omega}} d_{ij} | i \rangle \langle j|,
\end{equation}
where $d_{ij} = \langle i | (-\mathbf{ D }) | j \rangle / \hat{D}$ is the dimensionless value of the matrix element of the dipole moment operator $\mathbf{D}$ (dimensional unit $\hat{D}$). It is convenient to pass to the continuum limit of environmental modes $\nu_k$ such that $\sum_k |\alpha_k|^2 \to \int \textrm{d} \nu\, J(\nu )$.
Here, $\alpha_k$ describes the coupling strength between the system and the reservoir mode $\nu_k$,
and $J(\nu)$ is the spectral density of electromagnetic modes~\cite{Breuer2002}. Considering only the zero temperature environment, where all the modes are initially empty, each decay channel is related to a time-dependent decay rate
\begin{equation}
\Delta_{\omega} (t) = 2 \int_{0}^{t} \textrm{d}s \int_{0}^{\infty} \textrm{d}\nu\, J(\nu) \cos [(\nu - \omega ) s ].
\end{equation}

The interaction with the reservoir introduces a renormalization of the system Hamiltonian $H_S$ by a Hermitian term, i.e., the Lamb shift Hamiltonian
\begin{equation}
H_{LS}(t) = \hbar \sum_{\omega} \lambda_{\omega} (t) C_{\omega}^{\dagger} C_{\omega},
\end{equation}
where the time-dependent rate factor is
\begin{equation}
\lambda_{\omega} (t) = \int_{0}^{t} \textrm{d}s \int_{0}^{\infty} \textrm{d}\nu\, J(\nu) \sin [(\nu - \omega ) s ].
\end{equation}
We label the different Bohr frequencies by $\{ \omega^j \}$, where $j=1,\ldots$. Correspondingly, the jump operators are $C_j \equiv C_{\omega^j}$ and the decay rates are $\Delta_j (t) \equiv \Delta_{\omega^j} (t)$ and $\lambda_j (t) \equiv \lambda_{\omega^j} (t)$. Then, the time-local master equation in the interaction picture is in the form of Eq.~\eqref{Eq:MNM}, where system Hamiltonian $H_S$ has been replaced by $H_{LS} (t)$.

The spectral density of the electromagnetic field inside an imperfect cavity is well approximated by a Lorentzian distribution
\begin{equation}
J_{\textrm{Lorentz}} (\nu ) = \frac{\alpha^2}{2\pi} \frac{\Gamma}{(\nu - \omega_{\textrm{cav}} )^2 + (\Gamma / 2)^2},
\end{equation}
where $\alpha^2$ is a coupling constant, $\omega_{\textrm{cav}}$ is the resonance frequency of the cavity and $\Gamma$ characterizes the width of the distribution. The essential parameter in this case is the detuning $\delta_j \equiv \omega_{\textrm{cav}} - \omega^j$ of the Bohr frequency with respect to the cavity resonance frequency. 

Since the cavity supports only modes residing close to its resonance frequency $\omega_{\textrm{cav}}$, only transitions whose Bohr frequencies are close to this value contribute to the dynamics. This justifies the description of the atom's Hilbert space consisting effectively of only two or three levels, which we now study.

\subsection{Units and parameters}

In the examples, the time scale is set by the inverse of the spectral distribution width $\Gamma^{-1}$. The resonance frequency is assumed to be large $\omega_{\textrm{cav}} \gg \Gamma$. The Markovian time scale is then $\tau_{M} \sim 10\, \Gamma^{-1}$ [c.f. convergence of the decay rates to steady Markovian values in, e.g., Fig \ref{fig:JaynesCummingsResults}(a)]. In the Jaynes-Cummings model the coupling constant is set to $\alpha^2 = 5$ and in the three-level systems it is $\alpha^2 = 2$. The dipole moment matrix elements are always assumed to be $d_{ij} = 1$ for all pairs of states $i\neq j$. In the numerical simulations the time step size is $\delta t = 0.01 \Gamma^{-1}$ and the size of the ensemble is $N=10^5$. The notation of atomic levels is the same as in Fig.~\ref{fig:examplesSchematic}.

\subsection{\label{sec:exampleResults}Results}

For the sake of comparison, we solve the master equation in two different ways. First, we solve the density matrix by using the NMQJ method. Second, we calculate the formal analytical solutions of the equations of motion of the individual density matrix components (expressions are given in Appendix B). The results are then compared in order to verify the functionality of our method.

\subsubsection{Two-level atom: detuned Jaynes-Cummings model}\label{subsubsec:JC}

\begin{figure}[tb]
\includegraphics{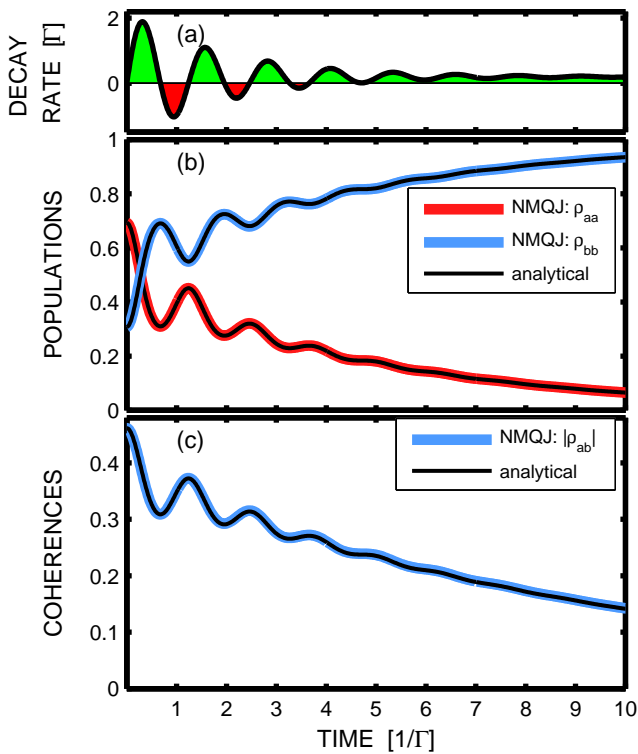}
\caption{\label{fig:JaynesCummingsResults}
(Color online) Dynamics of the Jaynes--Cummings model. Initial state is $|\psi_0 (0) \rangle = (3|a\rangle + 2 |b\rangle )/\sqrt{13}$. (a) Decay rate $\Delta_1 (t)$. (b) Populations $\rho_{aa}$ (initially higher line) and $\rho_{bb}$ (initially lower line). (c) Absolute value of the coherence $\rho_{ab}$.}
\end{figure}

The two-level case involves only one Lindblad operator $C_1 = \sigma_- = | b \rangle \langle a |$, which is the usual lowering operator from the excited to the ground state. We choose the detuning $\delta_1 = 5 \, \Gamma$ and the Fig.~\ref{fig:JaynesCummingsResults} (a) shows the oscillatory behavior of the corresponding decay rate $\Delta_1 (t)$ . The initial state is a pure state $\rho (0) = |\psi_0(0) \rangle \langle \psi_0 (0)|$, meaning that all the $N$ ensemble members are initially in the same state $|\psi_0 (0) \rangle$. In our example $|\psi_0 (0) \rangle = ( 3 |a\rangle + 2 |b\rangle )/\sqrt{13}$.

For the given single jump operator and an initial state including a finite excited state component, there will be only two kinds of states contributing to the master equation solution. This is because according to the unraveling in Eq.~\eqref{Eq:Rho} the global phase factors of the single ensemble members do not affect the density matrix representation. The two non-equivalent states are now the evolved initial state vector $|\psi_0 (t) \rangle$ and the ground state $|\psi_1 \rangle \equiv |b\rangle$, which can be reached from $|\psi_0 (t)\rangle$ by operating with the Lindblad operator. Correspondingly, there are two discrete variables $N_0 (t)$ and $N_1 (t)$ counting the number of ensemble members on each of these two states. Initially $N_0 (0) = N$ and $N_1 (0) = 0$.  

For a certain initial time interval, the decay rate $\Delta_1$ is positive (see Fig.~\ref{fig:JaynesCummingsResults}).
During this period the ensemble evolves according to the standard MCWF description. The deterministic evolution $|\psi_\alpha (t) \rangle \to |\psi_\alpha( t + \delta t) \rangle$ is given by Eq.~\eqref{eq:deterministicEvolution} with Hamiltonian $H = H_{LS} - \frac{i\hbar}{2} \Delta_1 (t) C_1^\dagger C_1 = \hbar [ \lambda_1 (t) - \frac{i}{2} \Delta_1 (t)] |a\rangle \langle a|$. The deterministic evolution is interrupted by quantum jumps $|\psi_0 (t) \rangle \to |\psi_1 (t) \rangle$ occurring with a probability $P_0^1 (t) = \Delta_1 (t) \delta t \langle \psi_0 (t)| C_1^\dagger C_1 |\psi_0 (t) \rangle = \Delta_1 (t) \delta t |\langle a |\psi_0 (t) \rangle |^2$ given by the Eq.~\eqref{eq:positiveJumpProbability}. In our notation this means that when quantum jump occurs, the occupation numbers are updated as $\{ N_0(t), N_1 (t) \} \to \{ N_0 (t) - 1, N_1 (t) + 1\}$. Once an ensemble member has jumped to the state $|\psi_1\rangle$, it can not experience any other quantum jumps during this period, since the corresponding jump probability is $P_1^1 \propto |\langle a| \psi_1 \rangle |^2 = 0$.

After the first positive period the decay rate becomes negative. The deterministic evolution is still driven by the same Hamiltonian as previously. However, now those ensemble members which had previously jumped to the ground state $|\psi_1\rangle$ are able to make a reverse non-Markovian quantum jump $|\psi_0 (t) \rangle \leftarrow |\psi_1 \rangle$ going back to the deterministically evolved initial state. The probability of this jump is given by Eq.~\eqref{Eq:JProb2M} and  is 
\begin{eqnarray}
P_{1\to 0}^1 (t) &=& \frac{N_0 (t) }{N_1 (t)} | \Delta_1 (t) | \delta t \langle \psi_0 (t) | C_1^\dagger C_1 | \psi_0 (t) \rangle \nonumber \\
&=& \frac{N_0 (t) }{N_1 (t)} | \Delta_1 (t) | \delta t |\langle a | \psi_0 (t) \rangle |^2.
\end{eqnarray}
Accordingly, the occupation numbers are updated after each reverse jump such that $\{ N_0(t), N_1 (t) \} \to \{ N_0 (t) + 1, N_1 (t) - 1\}$. The ensemble members in the state $|\psi_0 \rangle$ are not able to perform quantum jumps during this period, since in the ensemble there are no states $|\psi_\alpha\rangle$ for which $|\psi_0 (t) \rangle = C_1 |\psi_\alpha\rangle / \| C_1 |\psi_\alpha \rangle \|$.

\begin{figure}[tb]
\includegraphics{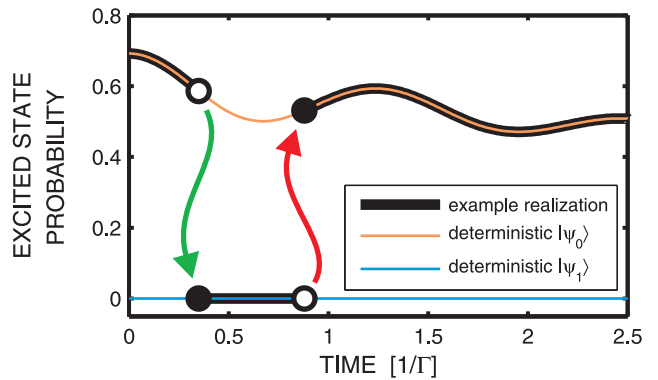}
\caption{\label{fig:JaynesCummingsExample} (Color online) Example of the dynamics of a single realization in the Jaynes-Cummings model (parameters as in Fig.~\ref{fig:JaynesCummingsResults}). In this realization the deterministic evolution of the initial state $|\psi_0\rangle$ (higher line) was followed until $t\approx 0.4 \, \Gamma^{-1}$, when a Markovian quantum jump (arrow down) brought the state to the ground state $|\psi_1\rangle = |b\rangle$ (lower line). Then, deterministic evolution of the ground state was followed until a non-Markovian quantum jump (arrow up) recreated the $|\psi_0\rangle$ state at $t \approx 0.8 \, \Gamma^{-1}$, whereafter the evolution was deterministic. }
\end{figure}

\begin{figure}[tb]
\includegraphics{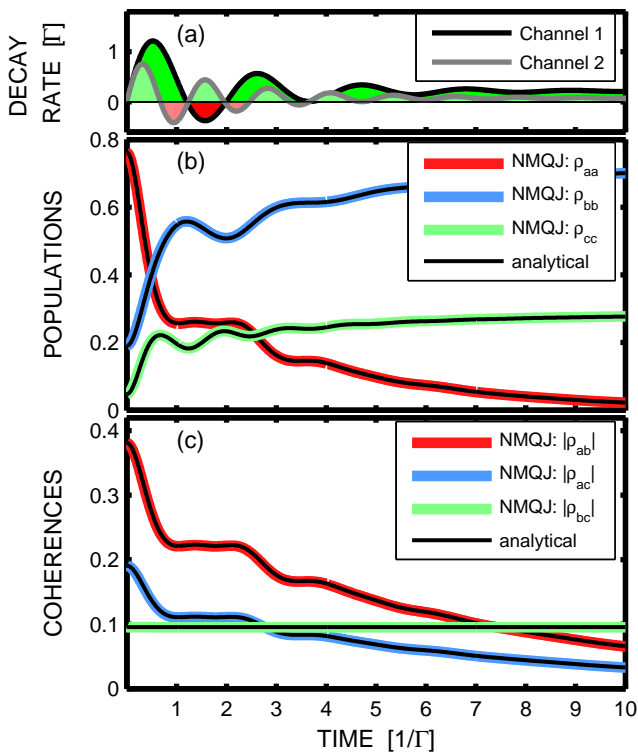}
\caption{\label{fig:lambdaResults} 
(Color online)
Dynamics of a $\Lambda$-system with an initial state $|\psi_0 (0) \rangle = (4 |a\rangle + 2 |b\rangle + |c\rangle ) /\sqrt{21}$. (a) Decay rates have momentarily opposite signs. (b) A plateau emerges to the excited state population $\rho_{aa}$ (initially highest line) as the decay channels counteract each other. Other populations $\rho_{bb}$ (initially middle line) and $\rho_{cc}$ (initially lowest line) behave according to the two separate decay channels. (c) Also coherences $\rho_{ab}$, $\rho_{ac}$, and $\rho_{cb}$ (initially highest, middle, and lowest line, respectively) have plateaus.}
\end{figure}

\begin{figure}[tb]
\includegraphics{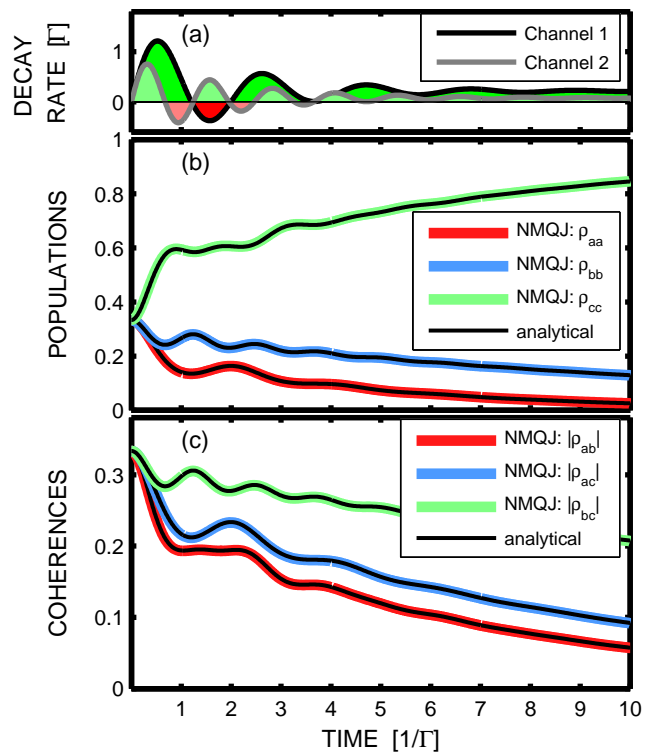}
\caption{\label{fig:veeResults} (Color online) Dynamics of a $V$-system with an initial state $|\psi_0 (0) \rangle = (|a\rangle + |b\rangle + |c\rangle ) /\sqrt{3}$. (a) Decay rates. (b) Populations of the upper states $\rho_{aa}$ and $\rho_{bb}$ (lowest and middle line, respectively) decay according to single decay channels and the ground state population $\rho_{cc}$ (highest line) increases correspondingly. (c) The upper state coherence $\rho_{ab}$ (lowest line) has a plateau as the decay channels are counteracting each other, while coherences involving the ground state $\rho_{ac}$ (middle line) and $\rho_{bc}$ (highest line) are related to individual decay channels.}
\end{figure}

The successive periods of positive and negative decay rate are treated in a similar way. In the Fig.~\ref{fig:JaynesCummingsResults} we show how the ensemble average of single realizations generated by the NMQJ method gives the exact solution of the master equation. In the corresponding Markovian case with a constant decay rate $\Delta_\textrm{Markov} = \lim_{t\to\infty} \Delta (t)$, the solution would be a simple exponential decay towards the ground state accompanied by exponential decoherence. 
The non-Markovian time-dependent decay rate leads to a slower or faster decay compared to the Markovian exponential one.
Furthermore, since the decay rate takes negative values, the decay process can be  partially reversed. This leads to a regain of excited state probability and re-coherence.

In Fig.~\ref{fig:JaynesCummingsExample} we give an example of a single realization experiencing both a quantum jump to the ground state $|\psi_0\rangle \to |\psi_1\rangle$ during the positive decay rate and a reverse non-Markovian quantum jump back to the initial state $|\psi_0\rangle \leftarrow |\psi_1\rangle$, during the negative decay rate. The essence of this illustration is that after these two jumps the state is (up to an irrelevant global phase factor) precisely the same as if the evolution would have been purely deterministic. However, when evaluating the time evolution with the ensemble average, the total contribution of this realization is different from the contribution given by a realization with no jumps.

\subsubsection{Three-level atom: $\Lambda$-system}

In a $\Lambda$-system there are two jump channels with Lindblad operators $C_1 = |b\rangle \langle a |$ and $C_2 = |c\rangle \langle a |$. In our example we choose the corresponding detunings to be $\delta_1 = -3 \, \Gamma$ and $\delta_2 = 5 \, \Gamma$. With these values the two decay rates have at certain time intervals opposite signs [c.f.~Fig.~\ref{fig:lambdaResults} (a)]. We now look at the initial state  $|\psi_0 (0) \rangle = (4|a\rangle + 2 |b\rangle + |c\rangle)/\sqrt{21}$. 

Starting with such an initial state the ensemble consists of effectively three different states: $|\psi_0 (t) \rangle$, $|\psi_1 \rangle \equiv |b\rangle$, and $|\psi_2 \rangle \equiv |c\rangle$. There are now two competing processes affecting the time evolution of the initial state. Initially both decay rates are positive, but at $t\approx 0.5 \, \Gamma^{-1}$, channel 2 becomes negative. This means that after this moment, on the one hand, there are still quantum jumps through channel 1 away from the initial state $|\psi_0\rangle \to |\psi_1\rangle$, but on the other hand, channel 2 repumps the ensemble members back to the initial state by non-Markovian quantum jumps $|\psi_0\rangle \leftarrow |\psi_2\rangle$. At $t\approx 1.2 \, \Gamma^{-1}$ both decay rates change their signs and so on all the way until $t\approx 2.5 \, \Gamma^{-1}$. Fig.~\ref{fig:lambdaResults} illustrates that when the decay rates are counteracting each other, plateaus in the evolution of the density matrix elements can be observed.

\subsubsection{Three-level atom: V-system}

In the case of a V-system, the two jump channels are  $C_1 = |c\rangle \langle a |$ and $C_2 = |c\rangle \langle b |$. We choose the detunings as earlier: $\delta_1 = -3 \, \Gamma$ and $\delta_2 = 5 \, \Gamma$. We consider the initial state 
$|\psi_0 (0) \rangle = (|a\rangle + |b\rangle + |c\rangle)/\sqrt{3}$. In this case, the ensemble consists of effectively only two different states, since both Lindblad operators act as $|\psi_0 (t) \rangle \to |\psi_1\rangle \equiv |c\rangle$. 

The dynamics in Fig.~\ref{fig:veeResults} shows how the upper state probabilities decay according to the individual decay channels.
Since only $|\psi_0\rangle $ carries coherences, there is a plateau in the upper state coherences, as it is affected simultaneously by decoherence and recoherence.

\subsubsection{Three-level atom: Ladder-system}

The ladder-system induces the most complicated dynamics of the three three-level atomic schemes considered here. The Lindblad operators form a short cascade, $C_1 = |b\rangle \langle a|$ and $C_2 = |c\rangle \langle b|$,
so that the target state of the upper channel  can still decay further by another quantum jump. There are three different possible quantum jump processes: $|\psi_0 (t) \rangle \to |\psi_1 \rangle \equiv |b\rangle$ through channel 1, $|\psi_0 (t) \rangle \to |\psi_2 \rangle \equiv |c\rangle$ through channel 2, and $|\psi_1 \rangle \to |\psi_2 \rangle$ through channel 2. Therefore, the effective ensemble consist of three state vectors. 

\begin{figure}[tb]
\includegraphics{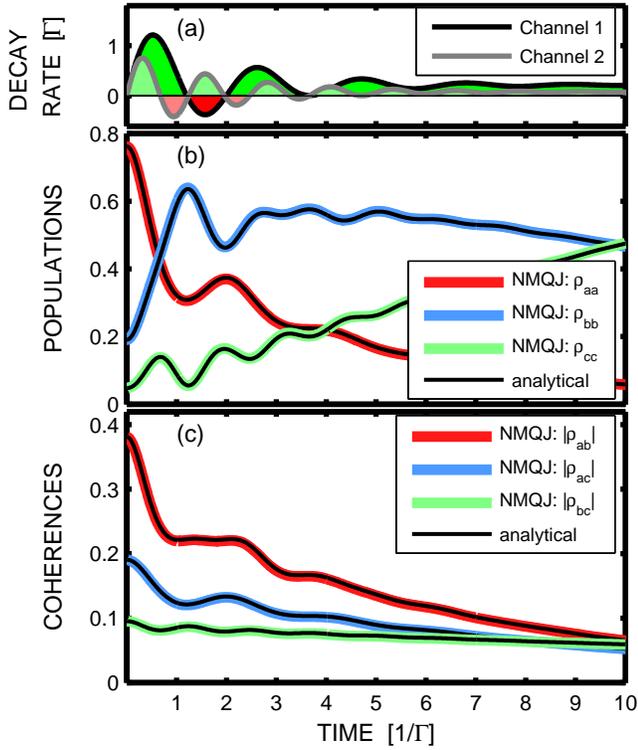}
\caption{\label{fig:ladderResults} 
(Color online)
Dynamics of a ladder-system with an initial state $|\psi_0 (0) \rangle = (4 |a\rangle + 2 |b\rangle + |c\rangle ) /\sqrt{21}$. (a) Decay rates. (b) The ladder structure is clearly visible as the decaying population of the highest excited state $\rho_{aa}$ appears first as increase of the middle state population $\rho_{bb}$, and eventually everything ends up to the ground state $\rho_{cc}$ (initially highest, middle and lowest line, respectively). (c) Only the initial state $|\psi_0\rangle$ contributes to the coherences $\rho_{ab}$, $\rho_{ac}$, and $\rho_{bc}$ (initially highest, middle, and lowest line, respectively), and therefore their dynamics is as simple as in the V-system.}
\end{figure}

During the negative period of channel 2, there are now interestingly two possible target states for a non-Markovian quantum jump from state $|\psi_2\rangle$ corresponding to processes $|\psi_0 (t) \rangle \leftarrow |\psi_2 \rangle$ and $|\psi_1 \rangle \leftarrow |\psi_2 \rangle$. The example dynamics in Fig.~\ref{fig:ladderResults} shows how the initial state $|\psi_0 (0) \rangle = (4 |a\rangle + 2 |b\rangle + |c\rangle ) /\sqrt{21}$ evolves. It is evident, that eventually the state decays towards $|\psi_2\rangle$, but due to complicated connections between the states and the changing signs of the decay rates, the dynamics is more rich than in the other cases.

\begin{figure}[tb]
\includegraphics{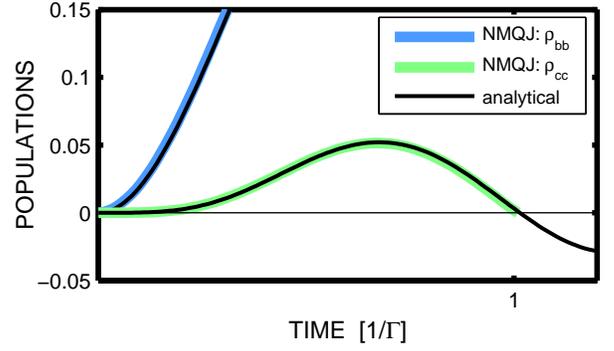}
\caption{\label{fig:ladderFailure} 
(Color online)
Dynamics of a ladder-system starting from an initial state $|\psi_0 (0) \rangle = |a\rangle$; other parameters are as in Fig.~\ref{fig:ladderResults}. Populations of the middle state $\rho_{bb}$ (higher line) and the ground state $\rho_{cc}$ (lower line) are shown. The NMQJ solution follows the formal analytical solution of the master equation faithfully until the state loses its positivity at $t\approx 1.0 \, \Gamma^{-1}$ as the ground state populations tends negative (thin black line marks the zero value). }
\end{figure}

Starting from an initial state $|\psi_0 (0) \rangle = |a\rangle$, our other example of ladder-system dynamics shows that the density matrix loses its positivity at $t \approx 1.0 \, \Gamma^{-1}$ (c.f. Fig.~\ref{fig:ladderFailure}), which indicates that the approximations in the derivation of the master equation do not hold for this level geometry.  The NMQJ solution follows the formal mathematical solution as long as it remains positive and the method is able to identify the point where the time evolution becomes unphysical. The failure of the positivity occurs, when channel 2 is still negative while all the ensemble members in state $|\psi_2\rangle$ have already had a non-Markovian quantum jump to states $|\psi_0 (t) \rangle$ and $|\psi_1 \rangle$. This happens, because the probability for such a non-Markovian quantum jump is $P_{2\to \alpha}^2 \propto N_\alpha(t) / N_2 (t)$, where $N_2 (t) \to 0$. This property has some interesting implications in the search for a positivity conditions for non-Markovian systems~\cite{Breuer08a}.

\section{Discussion}\label{Discu}

\subsection{On non-Markovian quantum jump operators and probabilities}\label{DisJOp}

To circumvent the problem of the negative probabilities of the Markovian MCWF method,  one is tempted to consider negative probabilities as positive ones for inverted jumps, i.e., to switch the role of the initial and final states of a given Lindblad operator by setting
$C_j \rightarrow C_j^{\dag}$. However, this does not lead to the correct ensemble for the non-Markovian dynamics. The essence of the negativity of the decay rate is the reversal of the decoherence process, i.e., re-coherence, and partial cancellation of the decoherence which occured in the past. If one uses in the non-Markovian region with negative decay rates the substitution $C_j \rightarrow C_j^{\dag}$, this only replaces one decoherent process with another one. 

Let us illustrate this with the simple example we considered  in Sec.~\ref{subsubsec:JC}. 
Writing the equations of motion explicitly for the density matrix elements of a two-level system,  gives for the positive decay rate region,
\begin{eqnarray}
\dot{\rho}_{aa}&=&  -|\Delta| \rho_{aa}, \nonumber \\
\dot{\rho}_{bb} &=& |\Delta| \rho_{aa}, \nonumber \\
\dot{\rho}_{ab} &=& -\frac{1}{2}|\Delta|  \rho_{ab},
\label{Eq:RhoElPos}
\end{eqnarray}
and for the negative decay region
\begin{eqnarray}
\dot{\rho}_{aa}&=&  |\Delta| \rho_{aa}, \nonumber \\
\dot{\rho}_{bb} &=& -|\Delta| \rho_{aa}, \nonumber \\
\dot{\rho}_{ab} &=& \frac{1}{2}|\Delta|  \rho_{ab}.
\label{Eq:RhoEl}
\end{eqnarray}
Here, $a$ denotes the excited and $b$ the ground state of the two-level atom.
The first line of Eq.~(\ref{Eq:RhoEl}) shows that during the negative decay the excited state probability increases  and that this increase is directly proportional 
to the probability which the excited state already has. This is a counterintuitive feature since it means that the 
total rate of the process is proportional to the target state, and not to the source state as in the positive decay
region [c.f.~Eq.~(\ref{Eq:RhoElPos})]. 
The last line of Eq.~(\ref{Eq:RhoEl}) shows that the coherences increase during the negative decay.

If one attempts to remedy the negative probability of the jump given by the Markovian method by changing the sign of the decay rate and substituting $C_j \rightarrow C_j^{\dag}$, or $\sigma_-\rightarrow \sigma_+$, this gives the equations of motion
\begin{eqnarray}
\dot{\rho}_{aa}&=&  |\Delta| \rho_{bb}, \nonumber \\
\dot{\rho}_{bb} &=& -|\Delta| \rho_{bb}, \nonumber \\
\dot{\rho}_{ab} &=& -\frac{1}{2}|\Delta|  \rho_{ab}.
\end{eqnarray}
It is easy to see that these equations are not the correct equations of motion (\ref{Eq:RhoEl}). In particular, the proportionality of the state populations for $\rho_{aa}$ and $\rho_{bb}$ go wrong, and the coherences decrease while the correct equations (\ref{Eq:RhoEl}) show that they must increase.

Generally speaking, a simple sign change of the decay rate from positive to negative in the non-Markovian master equation (\ref{Eq:MNM2}) may seem {\it a priori} as a rather 
trivial problem to solve. However, as the simple example above illustrates, the sign change actually leads to a very complicated problem. The main source of the complication is that the non-Markovian jump operators, given by Eq.~(\ref{Eq:Jump2}), do not appear explicitly in the master equation to be solved, whereas in the Markovian case one can pick the jump operators directly from the dissipator of the master equation. 

It is also interesting to note that 
we can interpret the jump probability (\ref{Eq:JProb2M}) in the following way. The numerator
$N_{\alpha'}|\Delta_{j_-}(t)| \delta t  \langle \psi_{\alpha'}(t) | C_{j_-}^{\dagger}(t) C_{j_-}(t) |\psi_{\alpha'}(t)\rangle$
gives the cumulative non-Markovian quantum jump probability in the whole ensemble. This is then divided to those
$N_{\alpha}$ ensemble members $|\psi_{\alpha}\rangle$ who perform the jumps.

\subsection{Why local-in-time master equation can describe non-Markovian dynamics with memory?}

Two common ways to describe non-Markovian open system dynamics are the memory kernel master equations and the local-in-time master equations with time dependent decay rates~\cite{Breuer2002}. The former consists of an integro-differential master equation where the change of the system state at a given moment of time is given by the integral over the past evolution
according to a given memory kernel. The local-in-time master equations in turn are based on the microscopic system-resevoir interaction modeling leading to a differential equation of motion for the density matrix of the system, which is local-in-time.
The description of non-Markovian dynamics without the use of a memory kernel, as done with the local-in-time master equations, may seem at first sight counterintuitive.
Our NMQJ method sheds new light on this issue and shows explicitly how and where the memory appears in local-in-time master equations. 

Suppose now that we have a density matrix of the system $\rho(t)$ and the corresponding ensemble of state vectors
during the initial positive decay region. At each time step a certain small fraction of the state vectors may jump
to decay channel $m$ according to the Markovian MCWF scheme:
$|\psi'\rangle \rightarrow |\psi\rangle = C_m |\psi'\rangle / || C_m |\psi'\rangle||$.
It is important to note that $|\psi'\rangle$ contains the information what the state $|\psi\rangle$
was before the jump  $|\psi'\rangle \rightarrow |\psi\rangle$ took place,
and that the whole ensemble still includes both types of state vectors $|\psi'\rangle$ and $|\psi\rangle$.
Then the system enters into the negative decay rate region.
Here, as described in the previous two subsections, the jumps go into opposite direction from $|\psi\rangle$ to $|\psi'\rangle$, and the probability of this jump is given by  the target state $|\psi'\rangle$. In other words, the very state vector that contains information on the past state of 
$|\psi\rangle$ defines both the target state of the non-Markovian jump and the probability for this jump to occur. In this way the past affects the current evolution of the system~\cite{Note:ManyPsi}.

It is difficult to see from the density matrix description where the memory of the earlier state of the system is. However, according to the description above, when we look at the density matrix as an ensemble of state vectors and study the dynamics of the state vectors in terms of the jumps, we see explicitly how the ensemble  members carry memory of other ensemble members. This memory comes into play when the decay rate becomes negative. 
It is also important to note that if the number of ensemble members in the target state of the reverse jump becomes equal to zero, $N_{\alpha'}=0$, then the system has lost its memory, and consequently the reverse jump probability vanishes since it is directly proportional to $N_{\alpha'}$ [c.f.~Eq.~(\ref{Eq:JProb2M})].

\subsection{Is continuous measurement of environment allowed for non-Markovian systems?}

For Markovian open quantum systems, single Monte Carlo realizations have a measurement interpretation~\cite{Plenio98}. The environment is thought to be monitored in a continuous way, and the corresponding reduced system evolution, conditioned on the measurement outcome, constitutes a single pure state trajectory of the ensemble. The existence of a measurement scheme interpretation for non-Markovian trajectories has been recently under active debate. Di\'osi claims that, at least in principle, certain types of QSD trajectories can be interpreted as true pure state single system trajectories~\cite{Diosi08}. His idea is based on the assumption of availability of an infinite set of entangled von Neumann detectors. Wiseman and Gambetta question Di\'osi's claims and the existence of true pure state trajectories with the measurement scheme interpretation. Their argument is based on the notion that in Di\'osi's scheme one should actually measure also those von Neumann apparatuses which are yet to interact with the system~\cite{Gambetta08}. Due to the entanglement between the von Neumann apparatuses, the measurement induces noise turning the true pure state trajectories into mixed ones.

Though both works mentioned above deal with diffusion descriptions, it is interesting to note how our jump scheme
fits into the discussion. In the NMQJ method, the memory of one ensemble member is carried by other ensemble members. When a reverse non-Markovian jump for a given ensemble member occurs, this member returns to the state which it would have at this point of time, if the prior positive decay jump had not occurred. In the simple two-level atom example, the superposition which was lost earlier gets restored by the non-Markovian jump, and the information on the earlier state of the system 
returns from the environment to the system. 
The crucial point is that the information lost by the system to the environment in the initial positive decay region has to be still available to the system when the decay rate turns later on negative. If we measure the environment in a continuous way, we are extracting information from the environment - and indirectly on the system state. If this measurement is destructive, then the information is not available to the system anymore and the non-Markovian dynamics gets distorted. In the case of a two-level atom, the measurement of the photon in the environment destroys the photon, and the two-level atom can not get re-excited during the negative decay region.

In addition,  in the two-level atom example, the oscillations in the excited state probability arise due to virtual exchanges of excitations between the system and the reservoir~\cite{Breuer2002,Breuer99,BreuerGen}. Virtual processes can not be directly measured while they still affect the system dynamics. 
This fits to the insight that the NMQJ gives though in terms of virtual processes there is a subtle difference: instead of virtual exchange of photons between the two-level atom and the reservoir, we rather describe the oscillations in the excited state amplitude of the atom as 
destruction and restoration of the quantum superposition. This difference between the two descriptions arises because an absorption of the photon by the atom means a jump from the ground state to the excited state. This process, by definition, can not increase the coherences which is a key feature of non-Markovian systems in the negative decay region, as discussed in detail Sec.~\ref{DisJOp}.

\begin{figure}[tb]
\includegraphics{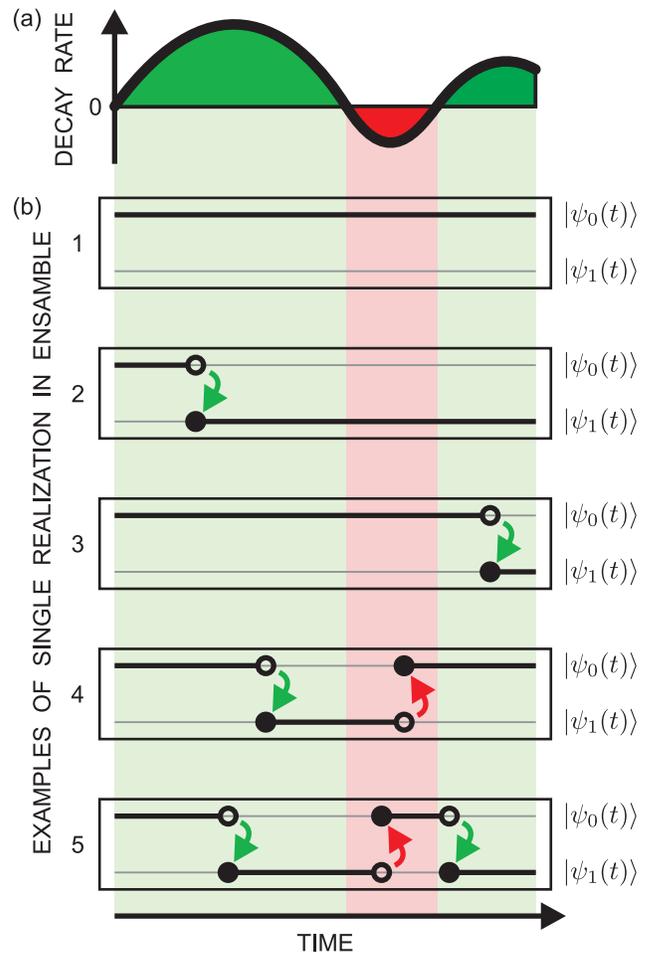}
\caption{\label{Fig:Paths}(Color online) (a) Sketch of a time-dependent decay rate with periods of positive and negative values (arbitrary units). (b) Examples of single realizations encountered in the ensemble. The system is assumed to be such that there is only one decay channel and two physically different states: the initial state $|\psi_0 (t) \rangle$ and the target state of a quantum jump $|\psi_1 (t) \rangle$ (deterministic evolution is given by thin horizontal lines). The state of an ensemble member at the given time is indicated by the thick line. Quantum jumps from $|\psi_0\rangle$ to $|\psi_1\rangle$ (arrows down) occur at random times during the positive decay rate, while non-Markovian quantum jumps from $|\psi_1\rangle$ back to $|\psi_0\rangle$ (arrows up) occur during the negative decay rate.
The total probability to have state $|\psi_0\rangle$ at the end of the shown evolution period is the sum of the paths 1 and 4 while the probability to have state $|\psi_1 \rangle$ is the sum of the paths 2, 3, and 5.
}
\end{figure}

If single realizations can not be measured, is there some other physical meaning that they have? In our formalism the probability to be in a given state at a given moment of time is the sum of all the paths leading to this state,
see Fig.~\ref{Fig:Paths}. In this sense the state vector evolutions can have an interpretation as possible paths that the system may take from its initial to final state. However, combining with the lack of measurement scheme, this means that we are not allowed to measure which path the system has taken while all possible paths contribute to the system state. If we try to extract information on the followed path by means of  measurements, we disturb the non-Markovian memory.
The rigorous connection to the Hilbert space path integral formalism will be studied in the future.

\subsection{Basic comparison to other jump descriptions}\label{Sec:Comp}

Earlier approaches to treat non-Markovian dynamics with quantum jumps 
use auxiliary states
and exploit the idea of Markovian embedding of non-Markovian dynamics in the extended Hilbert
space~\cite{Imamoglu,Garraway1997,Breuer99,BreuerGen}. 
Other jumplike unravelings use as an aid  the state of the total system and hidden variables~\cite{Gambetta2004} or take the measurement theory perspective~\cite{Collett}.
Our results show that it is possible to have jump-like unraveling of non-Markovian dynamics of the reduced system without extending the system Hilbert space or considering in detail the total system dynamics and hidden variables. It is worthwhile to see if the differences between our method and those developed earlier reveal interesting aspects of non-Markovian dynamics.
For this purpose, we compare our method to pseudomode (PM)~\cite{Garraway1997}, doubled~\cite{Breuer99} and triple~\cite{BreuerGen} Hilbert space methods (DHS and THS respectively), and to the quantum trajectory method based on hidden variables~\cite{Gambetta2004}.

The PM method describes the properties of the environment in terms of the  auxiliary pseudomode(s) with whom the system of interest
interacts~\cite{Garraway1997}. The pseudomode is then coupled to the Markovian reservoir while the system of interest interacts only, in a coherent way, with the pseudomode. The Markovian pseudomode master equation can be unravelled with the MCWF or some other Markovian method. Once this is done, the dynamics of the system of interest is obtained by tracing out the pseudomode.  This leads necessarily to mixed state trajectories for the system of interest while in our NMQJ method the time evolution of the ensemble members consists of pure states living in the Hilbert space of the system. 
In addition, the PM method relies on some assumptions on the form of the environment spectral density so that the pseudomode structure 
can be calculated, and it also exploits the solution of the total system dynamics. 
Our NMQJ method differs from the PM method in both of these issues and has been used to simulate two-level atom in photonic band gap in the absence of driving between the two states~\cite{Piilo08} (the driven case is more challenging, see the next subsection). 

On the other hand, the pseudomodes are by construction directly related to the properties of the environment.
As a matter of fact, it is possible to show by exploiting the insight given by the NMQJ method, that  the pseudomodes can be interpreted as an effective description of the memory of the environment of the open system~\cite{Mazzola08}. 
This is based on the notion that periods of negativity of the decay rate of local-in-time master equation coincide with those periods of time during which the pseudomode feeds coherently the system. 

The doubled Hilbert space (DHS) method uses two copies of the state vector to create a single realization in the ensemble~\cite{Breuer99}. 
The time evolution of the two copies is identical in the positive decay region. When the jumps with the Lindblad operators occur during the negative decay, 
one of the two copies gets multiplied by $-1$. This produces a negative contribution to the ensemble average. 
The probability in the ensemble is conserved because the norm of the deterministically evolving state vectors increases to values larger than one.
From the statistics point of view, this means that the number of jumps during negative decay has to match the increase of norm in the deterministic evolution, and the probability is conserved on average. The consequence is an additional source of statistical noise.
In the NMQJ method each state vector is normalized to one at each time step and the probability is conserved exactly. This gives a better statistical performance over the DHS method. In addition, the NMQJ  avoids the numerical burden which is present in the DHS method due to the doubling of the Hilbert space size~\cite{Erika}. 

An interesting improvement to the DHS method is provided by the triple Hilbert space (THS) method~\cite{BreuerGen}.
This method shows that the Markovian embedding 
of non-Markovian dynamics can be done with only three auxiliary discrete states. The original system dynamics is then contained in the coherences of the extended space state vectors. The method avoids the additional statistical noise term of the DHS method. However, the THS method uses a 4 times larger number of decay channels and a 3 times larger Hilbert space than the NMQJ method. Moreover, since the dynamics of the original system is contained in the coherences of the extended space,
unphysical situations such as violations of positivity of the density matrix during the time evolution may occur and pass unnoticed.
In contrast,  the NMQJ method, by construction, always keeps the dynamics positive since it is not possible to a have negative integer number of state vectors  in the ensemble. It is also worth mentioning that in the THS method the auxiliary quantum jump channels open when the decay rate becomes negative. This means that during negative decay interval, the probability flows out of the Hilbert space of the original system whereas in the NMQJ method, the direction of the probability flow within the Hilbert space of the system gets reversed at this point.

From the fundamental quantum physics point of view, it is also interesting to discuss the jumplike unraveling of non-Markovian dynamics which is based on hidden variables~\cite{Gambetta2004}. The basic idea of the method is to obtain the system trajectories from the guiding state describing the state of the total system. 
This is then used to obtain the stochastic evolution of the so called property state which includes information on the value of the environmental hidden variable and the corresponding properties of the reduced system. Our result seems to indicate that it is possible to describe non-Markovian dynamics with quantum jumps without the use of hidden variables. However, since the hidden variable approach allows jumps towards ground and excited states in the two-level atom case, it would be very interesting to compare in detail the time evolution of the ensemble members in both of the methods, and to see if there exists any connections between the two.

\subsection{Numerical and technical aspects}\label{Sec:Num}
Since in the NMQJ method the realizations depend on each other due to memory effects [c.f.~Eq.~(\ref{Eq:JProb2M})], it seems at first sight that all the $N$ ensemble members have to be evolved simultaneously.  However, according to Eq.~(\ref{Eq:Rho2}), the ensemble consists of several copies of each
$|\psi_{\alpha}(t)\rangle$. Obviously, there is no need to have on a computer several copies of the same state vector. It is sufficient to have one copy and the corresponding integer 
number $N_{\alpha}$.  Any number $N$ of the realizations of the process 
can be done by making $N_{\rm eff}\ll N$ state vector evolutions
where $N_{\rm eff}$ is equal to the number of terms in the summation $N=\sum_{\alpha}N_{\alpha}$ [c.f.~Eq.~(\ref{Eq:Rho2})].
When the realizations of the process are generated on a computer, a jump means changing the integer numbers $N_{\alpha}(t)$ accordingly in Eq.~(\ref{Eq:Rho2}). 
A saving in CPU time is achieved since it is not necessary at each point of time to evolve $N$ state vectors, instead, it is enough to decide $N$ times if the jumps occurred or not. This means that the NMQJ method has a built-in optimization which can be exploited to improve the efficiency of the method. For the two-level atom example described above, the effective ensemble size $N_{\rm eff}=2$ while $N=10^5$. However, these $N_{\rm eff}$ state vectors need to be evolved simultaneously since there is a dependence between the state vectors [c.f.~Eqs.~(\ref{Eq:Jump3}) - (\ref{Eq:JProb2M})]. 

We can summarize the key factors for the numerical performance of the NMQJ method as follows:
(i) no Hilbert space extensions are needed, (ii) the identification of the negative rate process as reverse jumps which keeps $N_{\rm eff}$ constant during the negative decay region and allows to technical optimization of the simulations (iii) the computational cost increases when the number of terms in the summation (\ref{Eq:Rho2}) increases. The first two points allow to improve the efficiency while the third point is expected to set the ultimate limit for the required computational resources. 
In addition of this resource limit, there exists also non-Markovian systems for which it is very challenging to derive local-in-time master equations of the form (\ref{Eq:MNM}). An example of this type of the system is a driven two-level atom in a photonic band gap material. To the best of our knowledge, there does not yet exist local-in-time master equations of the form (\ref{Eq:MNM}) for this system. On the other hand, it is possible to simulate this system already, e.g., with the method developed by Jack and Hope~\cite{JackPBG} which exploits memory functions and virtual density matrices.

In the quantum state diffusion (QSD) method~\cite{Strunz1999}, to obtain the operator giving the stochastic evolution of state vectors, one needs to perform a 
memory kernel integration combined with a functional derivative of the state vector with respect to the noise. In the NMQJ method the corresponding step goes in a fundamentally different way since the simulation produces its own non-Markovian quantum jump operator. This acts by transferring the ensemble members between the existing states in a stochastic way [c.f.~Eq.~(\ref{Eq:JOp})]. It is also worth mentioning that the QSD method by definition has continuous stochastic evolution of state vectors. This means that in the QSD simulation $N_{\rm eff} \sim N$.
For NMQJ method, when the complexity of the system to be treated increases, also $N_{\rm eff}$ increases. In the ultimate limit when the number of different state vectors is very large, or even approaches infinity, then there does not exist the optimization scheme for NMQJ method based on $N_{\rm eff}$.
In this case, the simulations also become more tedious due to the increasing number of state vectors which need to evolved simultaneously.

In general the derivation of local-in-time master equation for driven systems is a very challenging problem in the theory of non-Markovian open quantum systems. We believe that the main difficulties here are the condition of very strong driving affecting the system dynamics in the short non-Markovian time-scale, and the case of a very strong coupling between the system and the reservoir. In the latter case, the existence of a time-local generator of the reduced system dynamics is not in general guaranteed (see section 9.2.1
of Ref.~\cite{Breuer2002}). Hence, it is worth keeping in mind that the applicability of our method depends on this issue, since our starting point is the local-in-time master equation~(\ref{Eq:MNM}).

\section{Conclusions}\label{Conclu}

We have shown that, starting from a general local-in-time master equation, it is possible to describe the dynamics of a non-Markovian open system
with an ensemble of stochastic pure state evolutions with quantum jumps. The developed non-Markovian quantum jump method (NMQJ) demonstrates that it is indeed possible to unravel non-Markovian master equations with quantum jumps without making any auxiliary extensions to the Hilbert space of the system as done in the jump descriptions
developed earlier~\cite{Imamoglu,Garraway1997,Breuer99,Gambetta2004,BreuerGen}. Our approach allows a rather simple and insightful description of non-Markovian dynamics. Even though the method allows to optimize the simulations in terms of using the effective ensemble size $N_{\rm {eff}}$, this number increases with complexity of the system under study. This sets the limit for the performance of the method since $N_{\rm eff}$ state vectors need to be evolved simultaneously.

The NMQJ method developed here generalizes a widely used Markovian MCWF method~\cite{DCM1992} into the non-Markovian regime. Due to the existence of the negative decay rates for non-Markovian systems, the MCWF method leads to negative quantum jump probabilities. We have discovered the corresponding jump process which has positive probability. Due to the memory of the system, this non-Markovian quantum jump essentially acts as a reverse jump, and allows the system to recover the information lost earlier. 
The consequence is that in the ensemble of pure states forming the density matrix, the seemingly lost superpositions can be restored.
During the time evolution, jump -- reverse-jump cycles can occur in the ensemble members: the first jump during the positive decay destroys quantum superposition while the second jump in the negative decay region restores them. 

Our results shed new light on the non-Markovian dynamics in several ways. 
Breaking the density matrix evolution into an ensemble of state vectors 
with quantum jumps allows to understand how the density matrix carries the information on the earlier state of the system, and how the memory affects the system dynamics. This helps to clarify how local-in-time master equations are able to describe non-Markovian dynamics. Quantum mechanics reveals often counterintuitive features. Here,
the rate of the process appearing in the non-Markovian region is directly proportional to the target state of the process. This is opposite to the classical view where  typically the rate of a given process is given by the source state. Our analysis reveals in detail this counterintuitive feature of non-Markovian dynamics which is also present in the unravelled master equation.

It has been shown earlier that Markovian open system dynamics with MCWF trajectories can be formally described as a piecewise deterministic stochastic process of general probability theory~\cite{Breuer1995}. 
Consequently, we can ask what is the corresponding formal stochastic process for the NMQJ state evolutions~\cite{Breuer08a}.
This holds a promise to exploit new stochastic process  which may allow the ingredients and insight by our NMQJ method to be taken outside the field of open systems to a more general level. 

\acknowledgments
This work has been supported by the Academy of Finland
(Projects No.~108699, No.~115682, and No.~115982), the Magnus Ehrnrooth Foundation, the V\"ais\"al\"a Foundation, and the Turku Collegium of Science and Medicine. We thank H.-P. Breuer, B. Garraway, and J. Gambetta
for stimulating discussions.

\appendix

\section{}
In this Appendix A we show the details of the proof of the match between the master equation and the NMQJ method.

Averaging the evolution of the ensemble 
\begin{equation}
\label{Eq:RhoA}
\rho(t) = \sum_\alpha \frac{N_\alpha(t)}{N} |\psi_\alpha(t)\rangle \langle \psi_\alpha(t)|,
\end{equation}
over time step $\delta t$ gives
\begin{widetext}
\begin{eqnarray}
\label{Eq:Algo}
\overline{\sigma(t+\delta t)} &=&
\sum_{\alpha}\frac{N_{\alpha}(t)}{N} 
\left[
\left(
1-\sum_{j_+}P_{\alpha}^{j_+}(t) -\sum_{j_-,\alpha'} P_{\alpha\rightarrow \alpha'}^{j_-}(t)\right)
\right.
\frac{| \phi_{\alpha}(t+\delta t)\rangle \langle \phi_{\alpha}(t+\delta t) |}{||| \phi_{\alpha}(t+\delta t)\rangle ||^2}
\nonumber \\
&+&
\sum_{j_+} P_\alpha^{j_+}(t)
\frac{C_{j_+}(t) |\psi_{\alpha}(t)\rangle \langle \psi_{\alpha}(t)| C_{j_+}^{\dagger}(t) }
{||C_{j_+}(t) |\psi_{\alpha}(t)\rangle||^2}
+
\left.
\sum_{j_-,\alpha'} P_{\alpha\rightarrow \alpha'}^{j_-}(t) D_{\alpha\rightarrow \alpha'}^{j_-}(t) |\psi_{\alpha}(t)\rangle \langle \psi_{\alpha}(t)| D_{\alpha\rightarrow {\alpha}'}^{j_-\dagger}(t)
\right],\nonumber \\
\end{eqnarray}
\end{widetext}
where we have weighted, as usual, the deterministic evolution with the no-jump probability  
and the jump paths with the corresponding jump probabilities. Above,  we 
have the following quantities:
$P_{\alpha}^{j_+}(t)$ is the jump probability of the state $|\psi_{\alpha}(t)\rangle$ for positive channel $j_+$
\begin{equation}
\label{Eq:P+}
P_{\alpha}^{j_+}(t)=\Delta_{j_+} (t)\delta t \langle \psi_{\alpha}(t) | C_{j_+}^{\dagger}(t)C_{j_+}(t)|\psi_{\alpha}(t)\rangle,
\end{equation}
$P_{\alpha\rightarrow \alpha'}^{j_-}(t)$  is the reverse jump probability of state $|\psi_{\alpha}(t)\rangle$ via the negative channel $j_-$ to the state $|\psi_{\alpha'}(t)\rangle$
\begin{eqnarray}
P_{\alpha\rightarrow \alpha'}^{j_-}(t) &=& \frac{N_{\alpha'}(t)}  {N_{\alpha}(t)} |\Delta_{j_-}(t)| \delta t  
\nonumber \\
&\times&
\langle \psi_{\alpha'}(t) | C_{j_-}^{\dagger}(t) 
C_{j_-}(t) |\psi_{\alpha'}(t)\rangle.
\label{Eq:JProb2}
\end{eqnarray}
The reverse jump operator from the state $|\psi_{\alpha}\rangle = C_{j-}|\psi_{\alpha'}\rangle / ||C_{j-}|\psi_{\alpha'}||$ via channel $j_-$ to the state $|\psi_{\alpha'}\rangle$
is
\begin{equation}
D_{\alpha\rightarrow \alpha'}^{j_-}(t)= |\psi_{\alpha'}(t)\rangle \langle \psi_{\alpha}(t)|.
\label{Eq:JOpA}
\end{equation}

The deterministic evolution in Eq.~(\ref{Eq:Algo}) is given by
\begin{eqnarray}
\label{Eq:DetA}
| \phi_{\alpha}(t+\delta t)\rangle &=&
\left(1-\frac{i H_S\delta t}{\hbar} -\sum_j\frac{\Delta_j(t)\delta t}{2} C_j^{\dagger} (t)
C_j(t)\right) 
\nonumber \\
&\times&|\psi_{\alpha}(t)\rangle,
\end{eqnarray}
which gives for $| \phi_{\alpha}(t+\delta t)\rangle \langle  \phi_{\alpha}(t+\delta t) |$,
in first order in $\delta t$,
\begin{widetext}
\begin{eqnarray}
\label{Eq:DetAA}
| \phi_{\alpha}(t+\delta t)\rangle \langle  \phi_{\alpha}(t+\delta t) |
&=& |\psi_{\alpha}(t)\rangle \langle \psi_{\alpha}(t) | 
-i \delta t[H_S,  |\psi_{\alpha}(t)\rangle \langle \psi_{\alpha}(t) | ]
- \frac{\delta t}{2} \sum_j \Delta_j (t) \left\{C_j^{\dagger}(t) C_j(t), |\psi_{\alpha}(t)\rangle \langle \psi_{\alpha}(t) | \right\}.
\end{eqnarray}
\end{widetext}
It is easy to see from here, that this term gives the commutator and the anticommutator parts of the master equation.

In Eq.~(\ref{Eq:Algo}), the jump probabilities for the positive channel, appearing in the numerator and the denominator in the no-jump path, cancel each other when doing the series expansion in $\delta t$ and keeping the terms to first order. The jump part to positive channels gives the positive channel "sandwich term" of the master equation in the usual way.

We are left with the "sandwich" term for the negative channels.
Inserting  Eqs.~(\ref{Eq:P+})-(\ref{Eq:DetAA}) into  Eq.~(\ref{Eq:Algo}) and comparing to Eq.~(\ref{Eq:MNM2}) we have to show that
\begin{widetext}
\begin{eqnarray}
\label{Eq:LastStep}
&-& \sum_{\alpha, j_-} \frac{N_{\alpha}}{N}|\Delta_{j_-}(t)| \delta t
C_{j_-}(t) |\psi_{\alpha}(t)\rangle \langle \psi_{\alpha}(t)|C_{j_-}^{\dagger}(t) = 
\nonumber \\
&-&
\sum_{\alpha}\frac{N_{\alpha}}{N}  \sum_{\alpha',j_-} 
P_{\alpha \rightarrow \alpha'}^{j_-}(t)  |\psi_{\alpha}(t)\rangle \langle \psi_{\alpha}(t)| 
\delta\left( |\psi_{\alpha}(t) \rangle - \frac{C_{j_-}(t) |\psi_{\alpha'}(t)\rangle }{||C_{j_-}(t) |\psi_{\alpha'}(t)\rangle ||}\right)
\nonumber \\
&-& 
\sum_{\alpha}\frac{N_{\alpha}}{N}\sum_{j_-}
|\Delta_{j_-}(t)| 
\langle \psi_{\alpha}(t)| C_{j_-}^{\dagger}(t) C_{j_-}(t) |\psi_{\alpha}(t)\rangle \delta t
|\psi_{\alpha}(t)\rangle \langle \psi_\alpha (t) |
\nonumber \\
&+&
\sum_{\alpha}\frac{N_{\alpha}}{N}\sum_{j_-,\alpha'} 
P_{\alpha \rightarrow \alpha'}^{j_-}(t)  |\psi_{\alpha'}(t)\rangle \langle \psi_{\alpha'}(t)| 
\delta\left( |\psi_{\alpha}(t) \rangle - \frac{C_{j_-}(t) |\psi_{\alpha'}(t)\rangle }{||C_{j_-}(t) |\psi_{\alpha'}(t)\rangle ||}\right).
\end{eqnarray}
\end{widetext}
We have written here explicitly the $\delta$-functional which gives the condition for the reverse jump:
one can go via channel $j_-$ from $|\psi_{\alpha}\rangle$ to $|\psi_{\alpha'}\rangle$ on the condition that 
$|\psi_{\alpha}(t)\rangle =  C_{j_-}(t) |\psi_{\alpha'}(t)\rangle / ||C_{j_-}(t) |\psi_{\alpha'}(t)\rangle|| $.
In Eq.~(\ref{Eq:LastStep}), the last two lines cancel each other.  This happens because the $\delta$-functional takes care of the $\alpha$ summation in the last line and the summation over $\alpha$ and $\alpha'$ are equivalent procedures making the two terms equal with opposite signs. The first and second line in Eq.~(\ref{Eq:LastStep}) are equal. In the second line the $\delta$-functional with summation over $\alpha$ means  replacing $|\psi_{\alpha} \rangle$ with $C_{j_-} |\psi_{\alpha'}\rangle / \| C_{j_-} |\psi_{\alpha'}\rangle \|$ giving the sandwich term of the master equation in the first line.
Thus we have proven the equivalence between the master equation and the algorithm.

The proof can be summarized in the following way: the deterministic part gives the commutator and anticommutator parts of the master equation, the positive channels go in the usual way: the jump part giving the corresponding sandwich term of the master equation. For negative channels the change in the norm and jumps cancel and   the jump probability of negative channels times the deterministic evolution gives the sandwich terms.

\section{}

This appendix B gives the formal analytical solutions for the three-level systems considered in Sec. \ref{sec:exampleResults}. For simplicity, we neglect the Lamb-shift term.
First, let us define short-hand notation
\begin{align}
D_i (t) = \int_0^t \textrm{d} s \, \Delta_i (s), \\
L_i (t) = \int_0^t \textrm{d} s \, \lambda_i (s).
\end{align}
The direct formal solutions can be expressed by using these parameters, decay rates $\Delta_i (t)$, and initial conditions $\rho_{ij} (0)$ only.

\subsection*{Two-level atom: detuned Jaynes-Cummings model}

Master equation: 
\begin{align}
\dot \rho(t) = & \frac{1}{i} \lambda (t) [\sigma_+ \sigma_-, \rho(t) ] + \Delta (t) \sigma_- \rho(t) \sigma_+ \nonumber \\
& - \frac{1}{2} \Delta (t) \left\{ \rho(t), \sigma_+ \sigma_- \right \}.
\end{align}
Jump operator:
\begin{align}
C_1 &= \sigma_- = |b \rangle\langle a |.
\end{align}
Populations: 
\begin{align}
\rho_{aa} (t) &= e^{ -D_1(t)}\rho_{aa} (0), \\
\rho_{bb} (t) &= \Big\{ 1 - e^{ -D_1(t)} \Big\} \rho_{aa} (0) + \rho_{bb} (0).
\end{align}
Coherences:
\begin{align}
\rho_{ab} (t) &= e^{-D_1(t) / 2} \rho_{ab} (0).
\end{align}

\subsection*{Three-level atom: $\Lambda$-system}
Master equation: 
\begin{align}
\dot \rho(t) = & \frac{1}{i} \lambda_1 (t)[ |a\rangle\langle a|, \rho(t) ] + \frac{1}{i} \lambda_2 (t)[ |a\rangle\langle a|, \rho(t) ] \nonumber \\
& + \Delta_1 (t) \left[ | b \rangle\langle a | \rho(t) | a \rangle\langle b | - \frac{1}{2} \left\{ \rho(t), | a \rangle\langle a| \right \} \right] \nonumber \\
& + \Delta_2 (t) \left[ | c \rangle\langle a | \rho(t) | a \rangle\langle c | - \frac{1}{2} \left\{ \rho(t), | a \rangle\langle a| \right \} \right].
\end{align}
Jump operators:
\begin{align}
C_1 &= |b \rangle\langle a |,\\
C_2 &= |c \rangle\langle a |.
\end{align}
Populations:
\begin{align}
\rho_{aa} (t) &= e^{-[D_1 (t)+D_2(t)]} \rho_{aa} (0), \\
\rho_{bb} (t) &= \int_0^t \textrm{d} s \, \Delta_1 (s) e^{-[D_1 (s)+D_2(s)]} \rho_{aa} (0) \nonumber \\
& \quad + \rho_{bb} (0), \\
\rho_{cc} (t) &= \int_0^t \textrm{d} s \, \Delta_2 (s) e^{-[D_1 (s)+D_2(s)]} \rho_{aa} (0) \nonumber \\
& \quad + \rho_{cc} (0). 
\end{align}
Coherences:
\begin{align}
\rho_{ab} (t) &= e^{-[i L_1 (t) + i L_2 (t) + D_1 (t)/2 + D_2 (t)/2] } \rho_{ab} (0), \\
\rho_{ac} (t) &= e^{-[i L_1 (t) + i L_2 (t) + D_1 (t)/2 + D_2 (t)/2] } \rho_{ac} (0), \\
\rho_{bc} (t) &= \rho_{bc} (0).
\end{align}

\subsection*{Three-level atom: $V$-system}

Master equation: 
\begin{align}
\dot \rho(t) = & \frac{1}{i} \lambda_1 (t)[ |a\rangle\langle a|, \rho(t) ] + \frac{1}{i} \lambda_2 (t)[ |b\rangle\langle b|, \rho(t) ] \nonumber \\
& + \Delta_1 (t) \left[ | c \rangle\langle a | \rho(t) | a \rangle\langle c | - \frac{1}{2} \left\{ \rho(t), | a \rangle\langle a| \right \} \right] \nonumber \\
& + \Delta_2 (t) \left[ | c \rangle\langle b | \rho(t) | b \rangle\langle c | - \frac{1}{2} \left\{ \rho(t), | b \rangle\langle b| \right \} \right].
\end{align}
Jump operators:
\begin{align}
C_1 &= |c \rangle\langle a |,\\
C_2 &= |c \rangle\langle b |.
\end{align}
Populations: 
\begin{align}
\rho_{aa} (t) &= e^{-D_1 (t)} \rho_{aa} (0), \\
\rho_{bb} (t) &= e^{-D_2 (t)} \rho_{bb} (0), \\
\rho_{cc} (t) &= \Big[ 1 - e^{-D_1 (t)} \Big] \rho_{aa} (0) + \Big[ 1 - e^{-D_2 (t)} \Big] \rho_{bb} (0) \nonumber \\
& \quad + \rho_{cc} (0).
\end{align}
Coherences:
\begin{align}
\rho_{ab} (t) &= e^{-[i L_1 (t) + i L_2 (t) + D_1 (t)/2 + D_2 (t)/2] } \rho_{ab} (0), \\
\rho_{ac} (t) &= e^{-[i L_1 (t) + D_1 (t)/2] } \rho_{ac} (0), \\
\rho_{bc} (t) &= e^{-[i L_2 (t) + D_2 (t)/2] } \rho_{bc} (0).
\end{align}

\subsection*{Three-level atom: Ladder-system}

Master equation: 
\begin{align}
\dot \rho(t) = & \frac{1}{i} \lambda_1 (t)[ |a\rangle\langle a|, \rho(t) ] + \frac{1}{i} \lambda_2 (t)[ |b\rangle\langle b|, \rho(t) ] \nonumber \\
& + \Delta_1 (t) \left[ | b \rangle\langle a | \rho(t) | a \rangle\langle b | - \frac{1}{2} \left\{ \rho(t), | a \rangle\langle a| \right \} \right] \nonumber \\
& + \Delta_2 (t) \left[ | c \rangle\langle b | \rho(t) | b \rangle\langle c | - \frac{1}{2} \left\{ \rho(t), | b \rangle\langle b| \right \} \right].
\end{align}
Jump operators:
\begin{align}
C_1 &= |b \rangle\langle a |,\\
C_2 &= |c \rangle\langle b |.
\end{align}
Populations:
\begin{align}
\rho_{aa} (t) &= e^{-D_1 (t)} \rho_{aa} (0), \\
\rho_{bb} (t) &= e^{-D_2 (t)} \int_0^t \textrm{d} s \, \Delta_1 (s) e^{-D_1 (s)+D_2(s)} \rho_{aa} (0) \nonumber \\ 
&\quad + e^{-D_2 (t)} \rho_{bb} (0), \\
\rho_{cc} (t) &= \Big[ 1 - e^{-D_1 (t)} - e^{-D_2 (t)} \nonumber \\
&\quad \times \int_0^t \textrm{d} s \, \Delta_1 (s) e^{-D_1 (s)+D_2(s)} \Big] \rho_{aa} (0) \nonumber \\
&\quad + \Big[ 1 - e^{-D_2 (t) } \Big] \rho_{bb} (0) + \rho_{cc} (0). 
\end{align}
Coherences:
\begin{align}
\rho_{ab} (t) &= e^{-[i L_1 (t) - i L_2 (t) - D_1 (t)/2 - D_2 (t)/2] } \rho_{ab} (0), \\
\rho_{ac} (t) &= e^{-[i L_1 (t) + D_1 (t)/2 ] } \rho_{ac} (0), \\
\rho_{bc} (t) &= e^{-[i L_2 (t) + D_2 (t)/2 ] } \rho_{bc} (0).
\end{align}


\begin{thebibliography}{xx}
\bibitem{Note:ManyPsi}
Note that there can be several different paths and hence several different state vectors, as described in Fig.~3,
which contain information on the earlier state.
\end{thebibliography}
\end{document}